\documentclass[aps,prx,reprint,superscriptaddress]{revtex4-2}

\usepackage{graphicx}  

\usepackage{longtable}
\usepackage[T1]{fontenc}
\usepackage{dcolumn}   
\usepackage[inline]{enumitem}

\usepackage{bm}        
\usepackage{amsfonts}  
\usepackage{amsmath}   
\usepackage{amssymb}   
\usepackage[bb=boondox]{mathalfa}
\usepackage{color}
\usepackage[colorlinks=true,allcolors=blue]{hyperref}

\usepackage{floatrow}
\newcommand{\bra}[1]{\left\langle #1 \right|}

\newcommand{\pwisein}{\left\{ \begin{array}{ll}}
\newcommand{\pwiseout}{\end{array}\right.}
\newcommand{\ket}[1]{\left| #1 \right\rangle}

\newcommand{\trace}[1]{\mathrm{Tr} \left( #1 \right)}

\usepackage[caption=false]{subfig}

\usepackage{lipsum} 

\begin{document}

\title{Precision bounds for multiple currents in open quantum systems}

\author{Saulo V. Moreira}
\email{moreirsv@tcd.ie}
\affiliation{School of Physics, Trinity College Dublin, College Green, Dublin 2, D02 K8N4, Ireland}

\author{Marco Radaelli}
\affiliation{School of Physics, Trinity College Dublin, College Green, Dublin 2, D02 K8N4, Ireland}

\author{Alessandro Candeloro}
\affiliation{School of Physics, Trinity College Dublin, College Green, Dublin 2, D02 K8N4, Ireland}

\author{Felix C. Binder}
\affiliation{School of Physics, Trinity College Dublin, College Green, Dublin 2, D02 K8N4, Ireland}
\affiliation{Trinity Quantum Alliance, Unit 16, Trinity Technology and Enterprise Centre, Pearse Street, Dublin 2, D02 YN67, Ireland}

\author{Mark T. Mitchison}
\email{mark.mitchison@tcd.ie}
\affiliation{School of Physics, Trinity College Dublin, College Green, Dublin 2, D02 K8N4, Ireland}
\affiliation{Trinity Quantum Alliance, Unit 16, Trinity Technology and Enterprise Centre, Pearse Street, Dublin 2, D02 YN67, Ireland}

\begin{abstract} 

Thermodynamic (TUR) and kinetic (KUR) uncertainty relations are fundamental bounds constraining the fluctuations of current observables in classical, non-equilibrium systems. Several works have verified, however, violations of these classical bounds in open quantum systems, motivating the derivation of new quantum TURs and KURs that account for the role of quantum coherence. Here, we go one step further by deriving multidimensional KUR and TUR for multiple observables in open quantum systems undergoing Markovian dynamics. Our derivation exploits a multi-parameter metrology approach, in which the Fisher information matrix plays a central role.  Crucially, our bounds are tighter than previously derived quantum TURs and KURs for single observables, precisely because they incorporate correlations between multiple observables. We also find an intriguing quantum signature of correlations that is captured by the off-diagonal element of the Fisher information matrix, which vanishes for classical stochastic dynamics. By considering two examples, namely a coherently driven qubit system and the three-level maser, we demonstrate that the multidimensional quantum KUR bound can even be saturated when the observables are perfectly correlated.

\end{abstract}


\maketitle 

\section{Introduction}

One of the most consequential discoveries in the field of stochastic thermodynamics has been the interconnection between fluctuations of thermodynamic quantities and dissipation in non-equilibrium systems at the small scale.
Concepts from stochastic thermodynamics, such as stochastic entropy production~\cite{Seifert2005}, which corresponds to the entropy change of the system and reservoir due to their stochastic exchanges of heat within a given {\it trajectory}, have laid the foundation for a broader understanding of irreversibility and dissipation~\cite{Landi2021}. In turn, the number of stochastic transitions or jumps in a non-equilibrium system is related to time-symmetric frenetic aspects at the level of a single trajectory, with its average over a large number of trajectories, the {\it dynamical activity}, being a quantity which characterizes the volume of transitions in the system~\cite{Maes2017, DiTerlizzi2018, Maes2020}. 
Seminal findings such as detailed~\cite{Bochkov1981, Bochkov1981b, Crooks1998, Crooks1999, Seifert2004} and integral~\cite{Jarzyski1997, Jarzynski1997PRE} fluctuation theorems evidenced the interconnection between fluctuations of thermodynamic quantities and dissipation.
More recently, the discovery of precision bounds such as thermodynamic (TUR)~\cite{Barato2015} and kinetic (KUR)~\cite{DiTerlizzi2018} uncertainty relations, led to new foundational results in relation to the interplay between the {\it relative} fluctuation of observables and dissipation or dynamical activity in non-equilibrium systems. 

Both precision bounds, the TUR and KUR, were originally derived for classical systems and Markovian dynamics in the long time limit.
Specifically, the original TUR can be expressed as~\cite{Barato2015}
\begin{equation}\label{TUROG}
    \frac{{\rm Var} (\Theta)}{\langle \Theta \rangle^2} \ge \frac{2}{\Sigma},
\end{equation}
where $\Theta$ is a time-integrated current with mean $\langle \Theta \rangle$ and variance ${\rm Var}(\Theta)$, and $\Sigma$ is the system's entropy production (we set $k_B = 1$ throughout). In turn, the KUR is given by~\cite{DiTerlizzi2018}
\begin{equation}\label{KUROG}
    \frac{{\rm Var} (\Theta)}{\langle \Theta \rangle^2} \ge \frac{1}{\mathcal{A}},
\end{equation}
where $\mathcal{A}$ is the dynamical activity. These bounds imply that reducing the relative fluctuation of the current on the left-hand side of Eqs.~\eqref{TUROG} and~\eqref{KUROG} has a cost in terms of increased entropy production or dynamical activity. 

Soon after their original derivation using large-deviation theory~\cite{Gingrich2016,Garrahan2017}, it was realised that these precision bounds can also be obtained with tools from the theory of parameter estimation~\cite{Dechant2018, Hasegawa2019, Shiraishi2021}. In that context, observables, such as currents, are identified with estimators, and their fluctuations are interpreted as statistical errors in the estimate. The TUR and KUR can then be derived as a consequence of the Cram\'er-Rao bound, which imposes a fundamental lower limit on estimation error~\cite{Kay1993}. This approach has unlocked novel uncertainty relations for classical systems, encompassing multiple observables~\cite{Dechant2018, Dechant2021}, Langevin dynamics~\cite{VanVu2019,Dechant2020}, and first-passage times~\cite{Hiura2021,Pal2021b}, to name just a few examples. In many cases, TURs and KURs are obtained by imagining a fictional estimation of parameters representing a virtual (unphysical) perturbation to the dynamics, but nonequilibrium precision bounds can also be obtained by considering the estimation of physical parameters such as time~\cite{Prech2024}. 

Going beyond the scope of classical systems, significant efforts have been made to answer the question of whether classical bounds, such as those in Eqs.~\eqref{TUROG} and~\eqref{KUROG}, would be fulfilled by stochastic currents generated in non-equilibrium open quantum systems. In this context, several works have shown instances of violations of classical bounds by quantum dynamics described by Markovian quantum master equations in the long time limit, e.g. for quantum dot arrays~\cite{Ptaszynski2018, Agarwalla2018, Liu2019, Kacper2023}, the three-level maser~\cite{Kalaee2021}, and other nanoscale quantum engines~\cite{RignonBret2021}.  
Altogether, these results demonstrate that classical precision bounds do not generally constrain the relative fluctuation of observables in open quantum systems. Indeed, new bounds on the fluctuations of time-integrated currents or counting observables have been derived for quantum non-equilibrium steady states~\cite{Guarnieri2019} and for open quantum systems within the Markov approximation~\cite{Hasegawa2020, VanVu2022, Prech2024a}, which allow for reduced fluctuations compared to their classical counterparts. 
Notably, these results were all obtained using the Cram\'er-Rao bound of estimation theory, appropriately extended to open quantum systems. 

Despite this progress, the physical origin of quantum TUR violations remains an open question. To address this, recent works have studied the contribution of quantum coherence~\cite{Hasegawa2020,RignonBret2021,VanVu2022,Prech2024a} and entanglement~\cite{Kacper2023} to the fluctuations of a single counting observable. However, while certainly relevant, coherence and entanglement paint an incomplete picture, because they depend only on the quantum state. By contrast, current fluctuations depend on both the state and the dynamics~\cite{Landi2024}, e.g.~two distinct processes could yield the same nonequilibrium steady state but exhibit completely different current fluctuations. It is therefore natural to consider dynamical quantities, such as the correlations generated between sequentially measured observables. It is well known that these correlations may depart significantly from classical behavior due to quantum measurement invasiveness~\cite{LeggettGarg1985,Ruskov2006,Bednorz2012, Moreira2015,Knee2016,GarciaPintos2016}.

In this work, we take the first steps towards understanding how such correlations influence nonclassical current fluctuations far from equilibrium. Specifically, we derive bounds on the covariance of multiple counting observables that are measured simultaneously along the trajectories of an open quantum system. Our derivation exploits tools from  multi-parameter estimation theory, thus generalizing the results of Ref.~\cite{VanVu2022} to yield a multidimensional KUR and TUR. By capturing correlations between observables, our new bounds are tighter than previously derived TURs and KURs for quantum systems with Markovian dynamics. Our results are complementary to those of Ref.~\cite{Guarnieri2019}, which reported multidimensional precision bounds in the near-equilibrium regime without any Markovian assumption, whereas our bounds apply arbitrarily far from equilibrium so long as the Markov approximation is valid. Our results also reveal a novel quantum signature of correlations between pairs of simultaneously measured counting observables: namely, the off-diagonal element of the Fisher information matrix, which quantifies correlations between estimated parameters. We show that this quantity leads to an increased lower bound on the covariance of observables in generic quantum dynamics, whereas it vanishes for purely classical stochastic dynamics.


This article is organized as follows. In Section~\ref{Model}, we briefly present a general description of continuously monitored quantum systems. Section~\ref{Metrology} contains a brief overview of parameter estimation and the multi-parameter Cram\'er-Rao bound,  and
we review previously derived quantum KUR and TUR for single counting observables in Section~\ref{KURTUR}. In Section~\ref{MKUR/MTUR}, we present the multidimensional KUR and TUR derived in this work. Focussing on the multidimensional KUR, we illustrate the application of our results to two examples of non-equilibrium quantum systems, namely a coherently-driven qubit and the three-level maser system. Finally, we present our conclusions in Section~\ref{Conclusion}.



\section{Continuously monitored quantum systems}~\label{Model}
We consider a non-equilibrium quantum system described by the following Markovian quantum master equation~\cite{Lindblad1976, Gorini1976} (we set $\hbar = 1$ throughout), 
\begin{equation}\label{ME}
    \dot\rho = \mathcal{L}\rho = -i[H,\rho] + \sum_{k=1}^K\mathcal{D}[L_k]\rho,
\end{equation}
where $H$ is the system Hamiltonian, and $\mathcal{D}[L_k] \rho = L_k\rho L_k^\dagger - \frac{1}{2}\{L_k^\dagger L_k,\rho \}$ are dissipators, $L_k$ being jump operators.
The superoperator $\mathcal{L}$ is known as Liouvillian.
From a quantum trajectory perspective, Eq.~\eqref{ME} can be seen as the ensemble description of a quantum system continuously monitored by the environment~\cite{Plenio1998}.
In turn, monitoring a quantum system entails keeping a record of the observed jumps and their stochastic times. This gives rise to a quantum jump trajectory containing the jumps $k_j$ and their times $t_j$: $ \gamma \equiv \{ (t_1, k_1), (t_2, k_2), \dots (t_N, k_N) \}$, where $N$ is the (random) number of jumps that occur within the time interval $0 \leq t \leq \tau$. We assume the total time~$\tau$ of the evolution to be fixed.

This {\it unraveling} of the system dynamics into trajectories can be formalized using generalized measurements: at each infinitesimal time step $dt$, a measurement described by a quantum instrument is performed on the system, the Kraus operators of which are given by 
\begin{equation}\label{KrausOp}
    M_0 = 1 - i H_{\text{eff}} dt,  \ \
    M_k = \sqrt{dt}L_k,
\end{equation}
with $k = 1, 2, \dots, K$ and $H_{\text{eff}} \equiv H - \frac{i}{2}\sum_{k=1}^K  L_k^\dagger L_k $, which can be interpreted as an effective, but non-Hermitian, Hamiltonian.
The Kraus operators are normalized, satisfying $M_0^\dagger M_0 + \sum_k M_k^\dagger M_k = 1 + \mathcal{O}(dt^2)$, where $M_k$ represents a jump in channel $k$ and $M_0$ is associated with a no-jump event. The monitored system state $\rho_c(t)$ along a certain trajectory experiences the effect of Kraus operator $M_j$, $j=0,\ldots,K$, with probability $p_j^c=\trace{M_j\rho_c(t) M_j^\dagger}$, and is updated accordingly: $\rho_c(t+dt) = M_j \rho_c(t) M_j^\dagger/p_j^c$.

Given a pure initial system state $\ket{\psi_0}$, the probability of observing the trajectory $\gamma$ is given by
\begin{equation}\label{probgamma}
p(\gamma)=p_0|U(\tau,t_N)\Pi_{j=1}^NL_{k_j}U(t_j,t_{j-1})|\psi_0\rangle|^2,
\end{equation}
where $p_0$ is the probability associated with the preparation of $\ket{\psi_0}$, and $U(t_f, t_i) \equiv \exp[-i H_{\text{eff}} (t_f - t_i)]$ is a non-unitary, time propagation operator. 
 Thermal dissipation processes always present two jump channels for each bath, one with a reversed effect with respect to the other. We note that, generally, {\it local detailed-balance} is satisfied for these processes~\cite{Horowitz2013, Manzano2019, Manzano2022}. This means that the jump operators satisfy a time-reversal symmetry, i.e. $L_k = e^{\Delta s_k/2}L_{k^\prime}^\dagger$, where $L_{k^\prime}$ denotes the jump operator for reversed jump processes with respect to jumps represented by $L_k$, and $\Delta s_k$ is the environment entropy change due to the jump in channel $k$.

\section{Parameter estimation from quantum jump trajectories}\label{Metrology}

Suppose that we want to estimate an unknown parameter $\phi$, which is encoded on the system dynamics of a monitored open quantum system.
To tackle this problem, the framework for quantum parameter estimation from continuous measurements has been significantly developed in the last decades~\cite{Tsang2011, Gammelmark2014, Nurdin2022, Radaelli2023}. Within this framework,
from each trajectory $\gamma$, we can obtain an estimate for $\phi$ using an estimator $\Phi = \Phi(\gamma)$. Assuming that the estimator is unbiased, i.e. 
\begin{equation}
    \phi =\langle \Phi \rangle \equiv \sum_{\gamma} p_\phi(\gamma) \Phi (\gamma),
\end{equation}
where the expectation value is taken with respect to all the possible trajectories followed by the system, the following Cram\'er-Rao bound is satisfied~\cite{Fisher1922, Cramer1946}:
\begin{equation}\label{CRbound}
    \frac{{\rm Var}(\Phi)}{[\partial_\phi\langle \Phi \rangle]^2} \geq \frac{1}{F(\phi)}.
\end{equation}
Here, $F$ is the Fisher information, which is a function of the $\phi$-dependent probability distribution $ p_\phi(\gamma)$,
\begin{align}\label{FI}
    F(\phi) = \left \langle \left[\partial_\phi \ln p_\phi(\gamma)\right]^2\right\rangle = -\left\langle \partial_\phi^2 \ln p_\phi(\gamma) \right\rangle.
\end{align}
We remark that, in general, the Fisher information of a distribution over quantum trajectories is a complicated object to calculate. In this work, we used the monitoring operator formalism~\cite{Gammelmark2013,Albarelli2018,Radaelli2023}, summarised in Appendix~\ref{monitoring_operator_formalism}.

Consider, now, multiple unknown parameters $ \phi_1,  \phi_2,  \dots,   \phi_M $. Given $\vec{\phi} = \{\phi_1, \phi_2, \dots, \phi_M\}$, we denote the probability distribution depending on this set of parameters by $p_{\vec{\phi}}(\gamma)$. By considering a set of estimators $\vec{\Phi} = \{\Phi_1, \Phi_2, \dots, \Phi_M\}$, such that each $\Phi_i$ provides an estimate for a parameter $\phi_i$, the multi-parameter Cram\'er-Rao bound for the covariance matrix, $[\Xi(\vec{\Phi})]_{ij} \equiv \langle\Phi_i\Phi_j\rangle - \langle \Phi_i\rangle \langle \Phi_j \rangle$, can be expressed as the following matrix inequality~\cite{Kay1993, Dechant2018}:
\begin{equation}\label{matrixCR}
J_{\vec{\Phi}}^T \ \Xi^{-1}(\vec{\Phi}) \ J_{\vec{\Phi}}\leq\mathbb{F},
\end{equation}
where $ [J_{\vec{\Phi}}]_{ij} = \partial_{\phi_j}\langle \Phi_i \rangle $ is the Jacobian of $\langle \vec{\Phi} \rangle$ and the Fisher information matrix elements $[\mathbb{F}]_{ij} \equiv F_{ij}$ are given by
\begin{align}
   F_{ij} 
  &= \left\langle  \partial_{\phi_i} \ln p_{\vec{\phi}}(\gamma)\partial_{\phi_j} \ln p_{\vec{\phi}}(\gamma)\right\rangle \notag \\
  &  =  -\left\langle  \partial_{\phi_i} \partial_{\phi_j} \ln p_{\vec{\phi}}(\gamma)\right\rangle .
\end{align}


From the matrix inequality in Eq. \eqref{matrixCR}, we can introduce different scalar bounds, each representing a different optimality criterion \cite{Pukelsheim2006, Suzuki2021}. These generalise the single-parameter Cramér-Rao bound by including correlations between the parameters. One of the most popular criteria is D-optimality, which can be obtained from Eq.~\eqref{matrixCR} by taking the determinant of both sides of the inequality,
\begin{equation}\label{scalarCRbound}
    \frac{{\rm det(\Xi)}}{[\partial_{\phi_1}\langle\Phi_1\rangle]^2 \dots [\partial_{\phi_M}\langle\Phi_M\rangle]^2} \geq \dfrac{1}{{\rm det}(\mathbb{F})}.
\end{equation}
As we can see, this bound captures the covariances ${\rm Cov}(\Phi_i,\Phi_j) \equiv [\Xi]_{ij} $, for $i\neq j$, via the determinant of $\Xi$, thereby taking correlations between the observables $\Phi_i$ into account in Eq.~\eqref{scalarCRbound}. As a result, the latter bound is tighter than the product of the two single-parameter bounds given in Eq.~\eqref{CRbound}, with the gap precisely determined by the correlations between the observables $\Phi_i$.

\section{Reviewing the KUR and TUR in open quantum systems}~\label{KURTUR}

The quantum KUR and TUR in Ref.~\cite{VanVu2022} were derived for a single generic counting observable, which is defined as
\begin{equation}\label{countingobs}
    \Phi(\gamma) \equiv \sum_{j=1}^N w_{k_j},
\end{equation}
where $w_{k_j}$ is a stochastic variable that takes the value $w_{k}$ whenever a jump $j$ happens in channel $k$, with $w_k$ being arbitrary real coefficients associated with channel $k$.  For instance, when $w_k = -w_{k^\prime}$ ($k^\prime$ being the label of the reverse jump to $k$), $\Phi$ defines a time-integrated {\it thermodynamic current}. Note that we will use the terms time-integrated current and counting observable interchangeably throughout this paper, with the special case fulfilling $w_k = -w_{k^\prime}$ in Eq.~\eqref{countingobs} being denoted by $\Theta$ instead of $\Phi$.

Now, consider a perturbed system dynamics described by deformed $\phi$-dependent Hamiltonian, $H_\phi = (1 + \phi)H$, and $\phi$-dependent jump operators, $L_{k,\phi} = \sqrt{1 + \phi}L_k$ in Eq.~\eqref{ME}, with $\phi \ll 1$.
The probability distribution from Eq.~\eqref{probgamma} becomes dependent on $\phi$ and reads
\begin{equation}
    p_\phi(\gamma) = p_0(1+\phi)^{N}|U_{\phi}(\tau,t_N)\prod_{j}^N L_{k_{ j}}U_{\phi}(t_j,t_{j-1})| \psi_0 \rangle|^2,
\end{equation}
where $U_\phi(t_f, t_i) \equiv \exp[-i H_{\text{eff}, \phi} (t_f - t_i)]$ and $H_{\text{eff}, \phi} = (1 + \phi) H_{\text{eff}} $.
Note that $\phi = 0$ recovers the original dynamics. From a parameter estimation perspective, given the measurement record $\gamma$, $\Phi(\gamma)$ in Eq.~\eqref{countingobs} can be seen as an estimator for $\Psi(\phi = 0) = \langle \Phi \rangle$.
In this way, in order to obtain a quantum KUR associated with the {\it original} dynamics described by Eq.~\eqref{ME} from the Cram\'er-Rao bound in Eq.~\eqref{CRbound}, the Fisher information must also be evaluated at $\phi = 0$, which gives $F(0) = \mathcal{A} + \mathcal{Q}$. Here,
\begin{equation}
    \mathcal{A} \equiv \int_0^\tau dt \sum_{k=1}^K {\rm tr}\{L_k \rho(t) L_k^\dagger\},
\end{equation}
is the dynamical activity
and $\mathcal{Q}$ can be seen as contribution from the quantum evolution. In particular, $\mathcal{Q}$ can be shown to vanish for a purely classical master equation~\cite{DiTerlizzi2018,VanVu2022} (i.e.~$H=0$ and jump operators of the form $L_k = \sqrt{\gamma_{mn}}|m\rangle \langle n|$ for some fixed basis $\{\ket{n}\}$). Using the Cram\'er-Rao bound in Eq.~\eqref{CRbound}, the following KUR can be obtained~\cite{VanVu2022},
\begin{equation}\label{quantumKUR}
    \frac{{\rm Var}(\Phi)}{\langle \Phi \rangle^2} \ge \frac{(1+\varphi)^2}{\mathcal{A} + \mathcal{Q}} \equiv K,
\end{equation}
where $\varphi = \tau \partial_\tau \ln |\langle \Phi \rangle|/\tau| $ is a correction term that vanishes in the long-time limit, $\tau\to \infty$.

In turn, for the particular case of the time-integrated current $\Theta$ and by imposing detailed balance, and considering the following  parameter imprinting on the Hamiltonian, $H_\theta = (1+\theta)H$, and the jump operators, $L_{k,\theta} = \sqrt{1+l_k(t)\theta}L_k$,
where $l_k(t)$ is defined as
\begin{equation}\label{lkalphaM}
    l_{k}(t) \equiv \frac{{\rm tr}\{L_{k}\rho(t) L_{k}^\dagger\}- {\rm tr}\{L_{k^\prime}\rho(t)L_{k^\prime}^\dagger\}}{{\rm tr}\{L_{k}\rho(t)L_{k}^\dagger\} + {\rm tr}\{L_{k^\prime}\rho(t)L_{k^\prime}^\dagger\}},
\end{equation}
the following quantum TUR can be derived~\cite{VanVu2022},
\begin{equation}\label{quantumTUR} 
    \frac{{\rm Var}(\Theta)}{\langle \Theta \rangle^2} \ge \frac{2(1+\vartheta)^2}{\Sigma + 2\mathcal{Q}^\prime}.
\end{equation}
In Eq.~\eqref{quantumTUR}, $\Sigma \equiv \Delta S + \Delta S_E$ is the average entropy production, $\Delta S$ being the change in the entropy of the system and $\Delta S_E$ the change in entropy of the environment, which are given by
\begin{align}
    \Delta S = {\rm tr}\{\rho_0 \ln\rho_0  \} - {\rm tr}\{\rho_\tau \ln\rho_\tau  \},   \\
    \Delta S_E = \int_0^\tau dt \sum_{k=1}^K {\rm tr}\{L_k\rho(t)L_k^\dagger \}\Delta s_k,
\end{align}
where $\rho_0$ and $\rho_\tau$ are the system density matrices corresponding to the initial and final states, respectively.
The quantity $\mathcal{Q}^\prime$ is also associated with the quantum evolution, in the same sense as in Eq.~\eqref{quantumKUR}, and $\vartheta$ is a correction term~\cite{VanVu2022} whose form is not very insightful and so we do not quote it here.

In summary, beyond the corrections $\varphi$ and $\vartheta$, $\mathcal{Q}$ and $\mathcal{Q}^\prime$ also contribute to the bounds in Eqs.~\eqref{quantumKUR} and~\eqref{quantumTUR}, representing the genuinely quantum terms. 
$\mathcal{Q}$ and $\mathcal{Q}^\prime$, which are components of the Fisher information at $\phi = 0$, are essentially associated with quantum evolution between quantum jumps. More specifically, from the perspective of the unravelling of the dynamics into quantum trajectories
using the Kraus operators in Eq.~\eqref{KrausOp}, they can be seen as capturing information about the parameter encoded via the time propagation operator $U_\phi$ along each trajectory. 
Thus, the usefulness of the parameter estimation approach for the derivations of Eqs.~\eqref{quantumKUR} and~\eqref{quantumTUR} becomes evident: it is due to the fact that the statistics of the {\it no-jump} outcomes along the trajectories, corresponding to the Kraus operator $M_0$, now contribute to the Fisher information, therefore resulting in bounds that are suited to constrain the relative fluctuation of observables in open quantum systems. 

It is important to note, however, that there are no known bounds on the quantum correction terms $\mathcal{Q}$, $\mathcal{Q}'$ in the general case. Indeed, they may take either positive or negative values, in principle, where a positive value for $\mathcal{Q}$ indicates the potential for reduced fluctuations while a negative value indicates the converse. Therefore, the structure of the quantum KUR and TUR alone is not sufficient to identify from first principles when violations of the classical precision bounds occur.

We note that other approaches to deriving quantum KURs and TURs are possible using the theory of quantum parameter estimation~\cite{Hasegawa2020, Prech2024a}. There, precision is bounded by the \textit{quantum} Fisher information, which dictates the smallest possible estimation error achievable by any continuous measurement scheme, i.e.~via alternative stochastic unravellings of the master equation~\cite{Gammelmark2014}. However, these approaches generally lead to looser bounds (since the quantum Fisher information upper-bounds the classical one) so we do not consider them further here.

\section{KUR and TUR for multiple currents in open quantum systems}\label{MKUR/MTUR}


We now consider multiple general counting observables and derive multidimensional KUR and TUR for open quantum systems. In the following, we will denote by $S = \{k\}_{k=1}^K$ the set of all jump channels, and $\cup_\alpha S_\alpha = S$, with $S_\alpha \subseteq S$ corresponding to disjoint subsets of the jump channels.
Furthermore, the set of jumps $j$ in the channels belonging to $S_\alpha$ in a given trajectory will be denoted by $T_\alpha$, i.e.  $j \in T_\alpha$ if $k \in S_\alpha$. In this way, we can define multiple general counting observables as 
\begin{equation}\label{multicurrents}
    \Phi_\alpha(\gamma) 
\equiv \sum_{j \in T_\alpha} w_{ k_{j}},
\end{equation}
where $\omega_{k_j} = \omega_k$ for $k \in S_\alpha$, $\omega_k$ are the coefficients defining the counting observable $\Phi_\alpha$.  For example, $S_\alpha$ may comprise jumps driven by coupling to one particular bath, labelled by $\alpha$. Then, the counting observable $\Phi_\alpha$ counts only those jumps that occur due to the coupling with that bath, e.g.~$\Phi_\alpha$ may represent the particle or heat current flowing into bath $\alpha$.

\subsection{Multidimensional KUR}

To derive our multidimensional KUR bound for these multiple counting observables, we consider perturbing the system dynamics by imprinting the parameters $\vec{\phi} = \{\phi_1, \phi_2 \dots,\phi_M\}$. 
Specifically, we consider the following deformed Hamiltonian and jump operators
\begin{align}
    H_{\vec{\phi}} = \left(1 + \sum_{\alpha=1}^{M}\phi_\alpha\right)H, \label{phihamiltonianM} \\
    L_{k,\phi_\alpha} = \sqrt{1+\phi_\alpha} L_{k}, \quad k \in S_\alpha. \label{phijumpopM}
\end{align}
The dynamics described by Eqs.~\eqref{phihamiltonianM} and~\eqref{phijumpopM} can be seen as speeding up different parts of the original dynamics differently using multiple parameters $\phi_\alpha \ll 1$.



From here on, we consider only two counting observables to derive our precision bounds. In this case, the scalar multi-parameter Cram\'er-Rao bound~\eqref{scalarCRbound} reduces directly to a bound on the covariance ${\rm Cov}(\Phi_1,\Phi_2)$. Taking $\vec{\phi} = \{\phi_1, \phi_2 \}$ in Eq.~\eqref{scalarCRbound} and taking the limit $\phi_\alpha\to 0$, we derive the following multidimensional KUR (see Appendix~\ref{DerivationMKUR} for details),
\begin{widetext}
\begin{equation}\label{multidimensionalKUR}
    \frac{{\rm det}(\Xi)}{\langle \Phi_1\rangle^2\langle\Phi_2\rangle^2} = \frac{{\rm Var}(\Phi_1)}{\langle \Phi_1\rangle^2} \frac{{\rm Var(\Phi_2)}}{\langle\Phi_2\rangle^2} - \frac{\rm Cov(\Phi_1,\Phi_2)^2}{\langle \Phi_1\rangle^2\langle\Phi_2\rangle^2}  \geq  \frac{(1+\varphi_1)^2(1+\varphi_2)^2}{(\mathcal{A}_1+\mathcal{Q}_1)(\mathcal{A}_2+\mathcal{Q}_2)-F_{12}^2}\equiv K_{12},
\end{equation}
\end{widetext}
where $\mathcal{Q}_1$ and $\mathcal{Q}_2$ are related to the quantum evolution, being defined as $\mathcal{Q}_\alpha = - \langle \partial_{\phi_\alpha}^2 \ln q_{\vec{\phi}} \rangle|_{\vec{\phi}=\vec{0}}$, with $q_{\vec{\phi}} \equiv |U_{\phi_1,\phi_2}(\tau,t_N)\prod_{j=1}^NL_{k_{j}}U_{\phi_1,\phi_2}(t_j,t_{j-1})| \psi_0 \rangle|^2$. The off-diagonal elements of the Fisher information matrix are denoted by $F_{12} = F_{21}$, and the dynamical activities $\mathcal{A}_1$ and $\mathcal{A}_2$ are given by
\begin{equation}
     \mathcal{A}_\alpha \equiv \int_0^\tau dt \sum_{k \in S_\alpha} {\rm tr}\{L_k \rho(t) L_k^\dagger\}.
\end{equation}
By writing $\varphi_1 = \langle \Phi^\star_1 \rangle / \langle \Phi_1 \rangle $, we show in Appendix~\ref{DerivationMKUR} that, in the long time limit,
\begin{multline}\label{PhiStar}
    \langle \Phi^\star_1 \rangle \approx \\ - \frac{\tau}{2} \sum_{k \in S_2} \int_0^\infty dt \left\langle  [L_{k}^\dagger, W_1(t)] L_{k} + L_{k}^\dagger [W_1(t), L_{k}] \right\rangle_{\text{ss}},
\end{multline}
with 
$W_\alpha \equiv \sum_{k \in S_\alpha} L_{k}^\dagger L_{k} w_{k}$, and $\langle X \rangle_{\text{ss}} \equiv {\rm tr}\{ X \rho_{\text{ss}}\}$, $\rho_\text{ss}$ being the system's steady state. Similarly, the expression for $\langle\Phi_2^\star\rangle$ is obtained by replacing the index 1 by 2 in Eq.~\eqref{PhiStar}. Unlike for the single-observable KUR~\eqref{quantumKUR}, we generally have $\varphi_\alpha \neq 0$ even in the long-time limit.


As we can see, Eq.~\eqref{multidimensionalKUR}
 is a bound on the product of the relative fluctuations of both observables.
Note that the term in the denominator on the right-hand side of the inequality originates from the diagonal elements of the Fisher information matrix, $F_{\alpha\alpha} = \mathcal{A}_\alpha + \mathcal{Q}_\alpha$.
In relation to the off-diagonal element of the Fisher information matrix, $F_{12}$, we show in Appendix \ref{Offdiagonal} that, given that a single parameter is encoded on each jump operator, then $F_{12} = 0$ for any classical rate equation. Since this is the case in our Eq.~\eqref{phijumpopM}, we can conclude that $F_{12} > 0$ is necessarily due to parameter imprinting via the Hamiltonian in Eq.~\eqref{phihamiltonianM} and is, therefore, associated with the quantum evolution. We note that the correction terms $\mathcal{Q}_\alpha$ and $\varphi_\alpha$ can have either a positive or a negative sign and therefore may either increase or decrease the right-hand side of the bound. Conversely, the presence of a non-zero $F_{12}$ always increases the bound.

To compare our bound~\eqref{multidimensionalKUR} with the results of Ref.~\cite{VanVu2022} while accounting for correlations between counting observables, we derive a second multidimensional KUR by combining Ineq.~\eqref{quantumKUR} with the general bound ${\rm det}(\Xi) \ge 0$, which follows from the fact that the covariance matrix is positive semi-definite. We denote the correction terms in Eq.~\eqref{quantumKUR} for $\Phi_1$ and $\Phi_2$ by $\varphi$ and $\varphi^\prime$, respectively. The following bound can then be derived:
\begin{widetext}
\begin{equation}\label{multiKURVu}
    \frac{{\rm Var}(\Phi_1)}{\langle \Phi_1\rangle^2} \frac{{\rm Var(\Phi_2)}}{\langle\Phi_2\rangle^2}   - \frac{1}{2}\frac{\rm Cov(\Phi_1,\Phi_2)^2}{\langle \Phi_1\rangle^2\langle\Phi_2\rangle^2}\ge \frac{1}{2}\frac{(1+\varphi)^2(1+\varphi^\prime)^2}{(\mathcal{A} + \mathcal{Q})^2} \equiv \frac{1}{2} K_1 K_2.
\end{equation}
\end{widetext}
This follows simply by adding together the independent inequalities $\prod_{i=1}^2 {\rm Var}(\Phi_i)/\langle \Phi_i\rangle^2\geq K_1 K_2$ and $\prod_{i=1}^2 {\rm Var}(\Phi_i)/\langle \Phi_i\rangle^2\geq {\rm Cov}(\Phi_1,\Phi_2)/\langle \Phi_1\rangle^2\langle \Phi_2\rangle^2$. Since the multidimensional KUR in Eq.~\eqref{multiKURVu} results from the single-parameter Cram\'er-Rao bound in Eq.~\eqref{CRbound} when applied individually to each counting observable $\Phi_1$ and $\Phi_2$, it does not incorporate the off-diagonal element of the Fisher information matrix, $F_{12}$.

\subsection{Multidimensional TUR}
To derive the multidimensional TUR, we consider local detailed balance. Furthermore, we assume that all pairs of channels $k, k^\prime$ satisfying $L_k = e^{\Delta s_k/2}L_{k^\prime}^\dagger$, will belong to the same set $S_\alpha$.
We now consider thermodynamic current observables $\Theta_\alpha$ (corresponding to antisymmetric weights $w_k = -w_{k'}$) and the same parameter imprinting on the Hamiltonian as before, 
\begin{align}
    &H_{\vec{\theta}} = \left(1 + \sum_{\alpha=1}^{M}\theta_\alpha\right)H, \label{thetahamiltonianM} 
\end{align}    
but with a distinct parameter imprinting on the jump operators, 
\begin{equation}\label{thetajumpopM1}
    L_{k, \theta_\alpha} = \sqrt{1+ l_{k}(t)\theta_\alpha} L_{k}, \quad k \in S_\alpha, 
\end{equation}
where $l_k(t)$ is given in Eq.~\eqref{lkalphaM}. As before, we focus on only two current observables. In Appendix~\ref{DerivationMTUR}, we derive a multidimensional TUR by using the scalar Cram\'er-Rao bound in Eq.~\eqref{scalarCRbound} and the parameter imprinting in Eqs.~\eqref{thetahamiltonianM} and~\eqref{thetajumpopM1}, for $\vec{\theta} = \{\theta_1, \theta_2 \}=\vec{0}$,
\begin{widetext}
\begin{equation}\label{multidimensionalTUR}
       \frac{{\rm det}(\Xi)}{\langle \Theta_1\rangle^2\langle\Theta_2\rangle^2} = \frac{{\rm Var}(\Theta_1)}{\langle \Theta_1\rangle^2} \frac{{\rm Var(\Theta_2)}}{\langle\Theta_2\rangle^2} - \frac{\rm Cov(\Theta_1, \Theta_2)^2}{\langle \Theta_1\rangle^2\langle\Theta_2\rangle^2} \geq \frac{2(1+\vartheta_1)^2(1+\vartheta_2)^2}{(\Sigma_1+2\mathcal{Q}^\prime_1)(\Sigma_2+2\mathcal{Q}^\prime_2)-2 F_{12}^2}.
\end{equation}
\end{widetext}
where $\mathcal{Q}^\prime_1$ and $\mathcal{Q}^\prime_2$ are contributions associated with the quantum evolution, given by $\mathcal{Q}_\alpha = - \langle \partial_{\theta_\alpha}^2 \ln q_{\vec{\theta}} \rangle|_{\vec{\theta}=\vec{0}}$, with $q_{\vec{\theta}} \equiv |U_{\theta_1,\theta_2}(\tau,t_N)\prod_{j=1}^NL_{k_{j}}U_{\theta_1,\theta_2}(t_j,t_{j-1})| \psi_0 \rangle|^2$. The components of the entropy production $\Sigma_1$ and $\Sigma_2$, which satisfy $\Sigma \equiv \Sigma_1 + \Sigma_2$, are given by
\begin{equation}
    \Sigma_\alpha \equiv \int_0^\tau dt \sum_{k\in S_\alpha}{\rm tr}\{L_{k}\rho(t)(\Delta s_{k}L_{k}^\dagger - [L_{k}^\dagger,\ln\rho(t)]) \}.
\end{equation}
By expressing $\vartheta_1 = \langle \Theta_1^\star \rangle / \langle \Theta_1 \rangle$, we show in Appendix~\ref{DerivationMTUR} that, by starting at the steady state, i.e. $\rho(0) = \rho_{\text{ss}}$, 
\begin{multline}\label{ThetaStar}
    \langle \Theta_1^\star \rangle =  \\   - \frac{\tau}{2} \sum_{k \in S_2} l_{k}^{\text{ss}}\int_0^\infty dt \left\langle  [L_{k}^\dagger, W_1(t)] L_{k} + L_{k}^\dagger [W_1(t), L_{k}] \right\rangle_{\text{ss}}.
\end{multline}
where $l_k^{\text{ss}}$ is a time-independent constant obtained via Eq.~\eqref{lkalphaM} for the steady state.
We note that, starting at the steady state and by following the same steps as in Appendix~\ref{Offdiagonal}, it can be shown that, as before, the off-diagonal element of the Fisher information, $F_{12}$, in Eq.~\eqref{multidimensionalTUR}, also becomes zero for classical master equations.

It is worth pointing out that, in general, while the dynamical activities $\mathcal{A}_\alpha$ satisfy $\mathcal{A} = \mathcal{A}_1 + \mathcal{A}_2$,
the elements $\Sigma_\alpha$ cannot be interpreted as the entropy change of their corresponding baths, as they also contain a term proportional to the entropy change of the system. In essence, this difference is due to the fact that the diagonal elements of the Fisher information matrix are now lower bounded by the components of the entropy production, $F_{\alpha \alpha} \leq \Sigma_\alpha + \mathcal{Q}_\alpha^\prime$, while for the dynamical activities the equality $F_{\alpha \alpha} = \mathcal{A}_\alpha + \mathcal{Q}_\alpha$ holds (See Appendixes~\ref{DerivationMKUR} and~\ref{DerivationMTUR} for details).
Note that there are no known bounds for $\mathcal{Q}_\alpha$ and $\mathcal{Q}_\alpha^\prime$, just like there are no known bounds for $\mathcal{Q}$ and $\mathcal{Q}^\prime$ from Ref.~\cite{VanVu2022}.

 We also note that the general form of Eqs.~\eqref{multidimensionalKUR} and~\eqref{multidimensionalTUR}, which incorporate $\mathcal{Q}_\alpha$ or $\mathcal{Q}_\alpha^\prime$ and $F_{12}$, is not arbitrary. The reason for this is that the parameter imprintings on the jump operators must be the ones given by Eqs.~\eqref{phijumpopM} and~\eqref{thetajumpopM1} in order to recover the dynamical activities $\mathcal{A}_\alpha$ in Eq.~\eqref{multidimensionalKUR} and $\Sigma_\alpha$ in Eq.~\eqref{multidimensionalTUR}. Other bounds can nonetheless be derived using the same imprinting for the jump operators, but considering different parameter imprintings for the Hamiltonian. Those bounds will, however, have the same structure as our multidimensional quantum bounds in Eqs.~\eqref{multidimensionalKUR} and~\eqref{multidimensionalTUR}. For example, even if the parameters are not imprinted on the Hamiltonian at all, $\mathcal{Q}_\alpha$ or $\mathcal{Q}_{\alpha}^\prime$ will still be different from zero because the ``no-jump'' evolution $M_0$ still depends on the imprinted parameters through the non-Hermitian effective Hamiltonian (see Eq.~\eqref{KrausOp}).

Analogously to Ineq.~\eqref{multiKURVu}, a second multidimensional TUR can also be obtained here by combining Ineq.~\eqref{quantumTUR} and ${\rm det}(\Xi) \ge 0$. By denoting the correction terms in Ineq.~\eqref{quantumTUR} for $\Theta_1$ and $\Theta_2$ by $\vartheta$ and $\vartheta^\prime$, respectively, we get,
\begin{equation}\label{multiTURVu}
    \frac{{\rm Var}(\Theta_1)}{\langle \Theta_1\rangle^2} \frac{{\rm Var(\Theta_2)}}{\langle\Theta_2\rangle^2}   - \frac{1}{2}\frac{\rm Cov(\Theta_1,\Theta_2)^2}{\langle \Theta_1\rangle^2\langle\Theta_2\rangle^2}\ge \frac{2(1+\vartheta)^2(1+\vartheta^\prime)^2}{(\Sigma + 2\mathcal{Q}^\prime)^2}.
\end{equation}
As we noted for Eq.~\eqref{multiKURVu}, this bound does not include $F_{12}$, despite incorporating the covariance term.

\begin{figure}
\includegraphics[width=0.95\columnwidth]{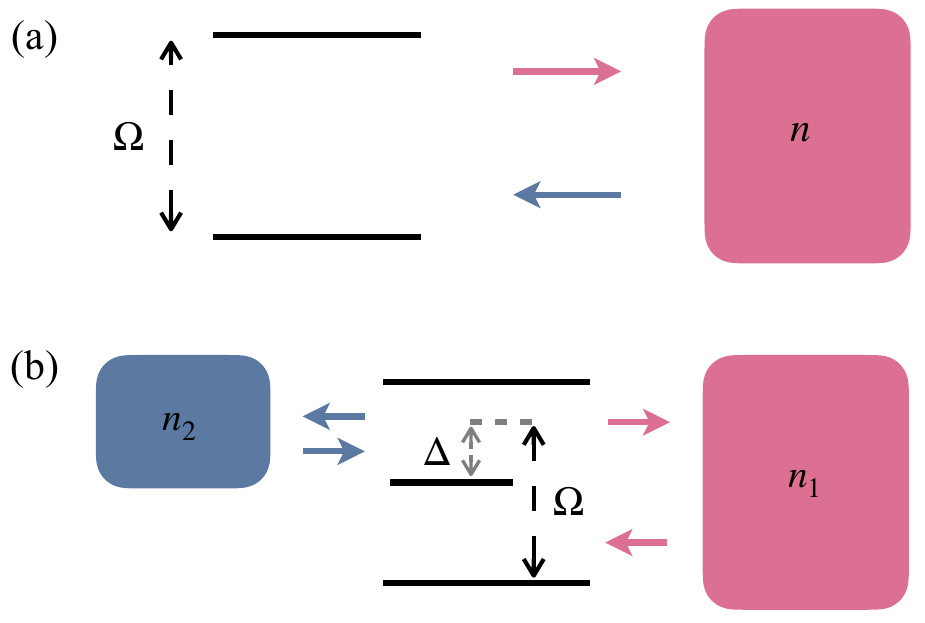}
\caption{(a) Sketch of a two-level system coherently-driven with strength $\Omega$ coupled to a bath with thermal occupation $n$. The arrows represent jumps in and out of the system, which we define as two distinct counting observables.   (b) Sketch of the three-level maser system, 
where $\Omega$ and $\Delta$ are the driving strength and the detuning, coupled to two thermal baths with occupations $n_1$ and $n_2$. Jumps in and out of each reservoir are represented with the same colour, defining different counting observables or heat currents.}\label{Fig1}
\end{figure}

\section{Examples}\label{Examples}

Focussing on the multidimensional quantum KUR, we now investigate two examples of nonequilibrium quantum systems, namely a coherently driven qubit and a three-level maser coupled to thermal baths, which are both sketched in Fig.~\ref{Fig1}.

\subsection{Coherently driven qubit}

We consider a coherently driven qubit system coupled to a thermal bath at temperature $T$. The system Hamiltonian is given by $H(t) = \omega \sigma_+\sigma_- + \Omega (\sigma_+ e^{-i\omega_dt} + \sigma_- e^{i\omega_dt})$, where  $\sigma_l$, with $l=x, y, z, +, -$, are Pauli matrices, $\omega$ is the system's bare frequency, and $\omega_d$ is the frequency of the drive. The eigenbasis of $\sigma_z$ is denoted by $\{ \ket{0}, \ket{1}\}$.
In an appropriate rotating frame, the Hamiltonian can be written in a time-independent way,
\begin{equation}
    \tilde{H} = \frac{\Delta}{2} \sigma_z + \Omega \sigma_x,
\end{equation}
where  $\Delta = \omega - \omega_d$ and $\Omega$ are the detuning and driving strength, respectively. The dynamics of the system is described by the quantum master equation 
\begin{equation}
    \dot{\rho}(t) =-i[\tilde{H},\rho(t)] + \mathcal{D}[L_1]\rho(t) + \mathcal{D}[L_{2}]\rho(t).
\end{equation}
The jump operators  are given by $L_1 = \sqrt{\gamma n}\sigma_+$ and $L_2 = \sqrt{\gamma (n+1)}\sigma_-$, with $\gamma > 0$ being the coupling strength; $n = (e^{\beta\omega} -1)^{-1}$ is the Bose-Einstein distribution, with $\beta = 1/T$.

\begin{figure}
\includegraphics[width=0.9\columnwidth]{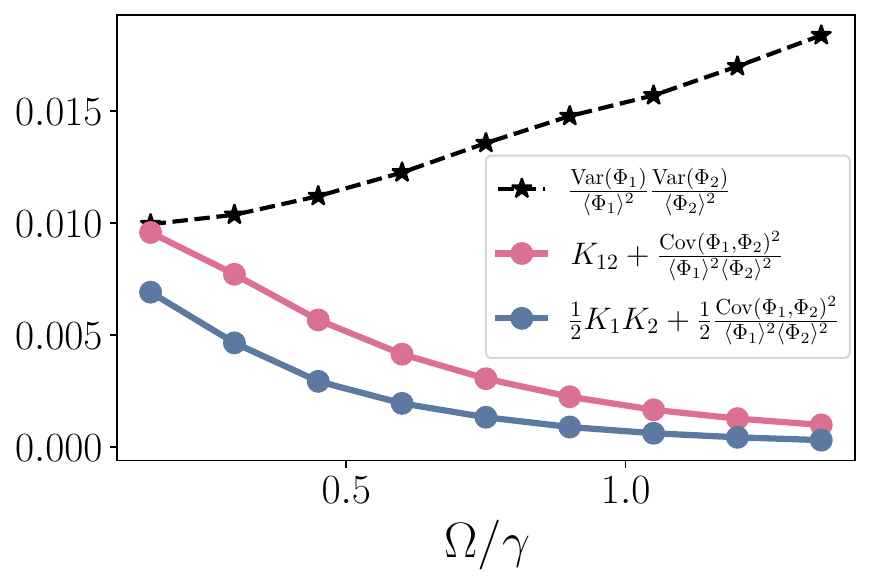}
\caption{\label{Fig2} 
 Plots of the product of the relative fluctuation of $\Phi_1$ and $\Phi_2$, and its bounds obtained from Eq.~\eqref{multidimensionalKUR}, which includes the term $K_{12}$, and from Eq.~\eqref{multiKURVu}, which includes $K_1 K_2$, as a function of $\Omega/\gamma$ for the coherently-driven qubit system.  The plots were obtained from the numerical simulation of $5 \times 10^4$ trajectories with $\tau = 10$.
The values of the parameters in both cases are $\Delta = 0$ and $n = 1$.}
\end{figure}

\begin{figure*}
\centering
\captionsetup[subfigure]{labelfont={large}}  
\sidesubfloat[]{%
\includegraphics[clip,width=0.43\columnwidth]{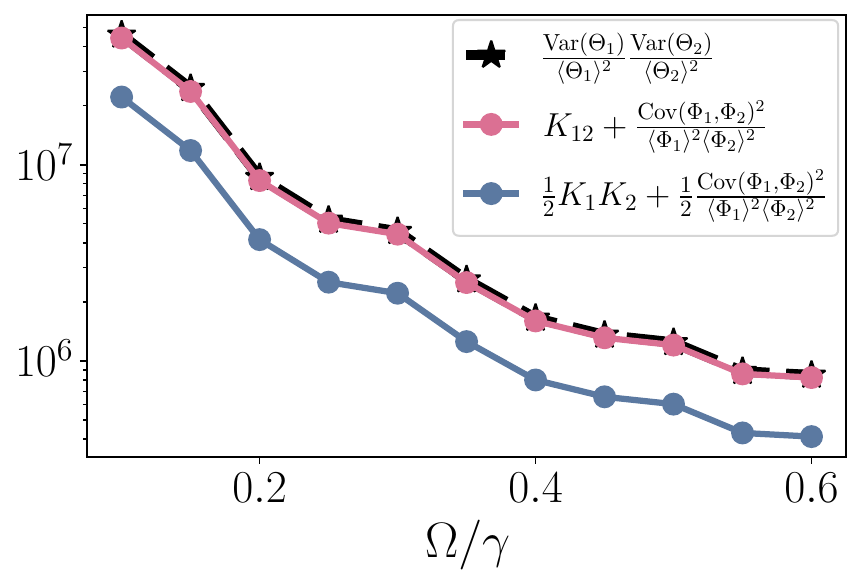}%
}
\hspace{1cm}
\sidesubfloat[]{%
\includegraphics[clip,width=0.43\columnwidth]{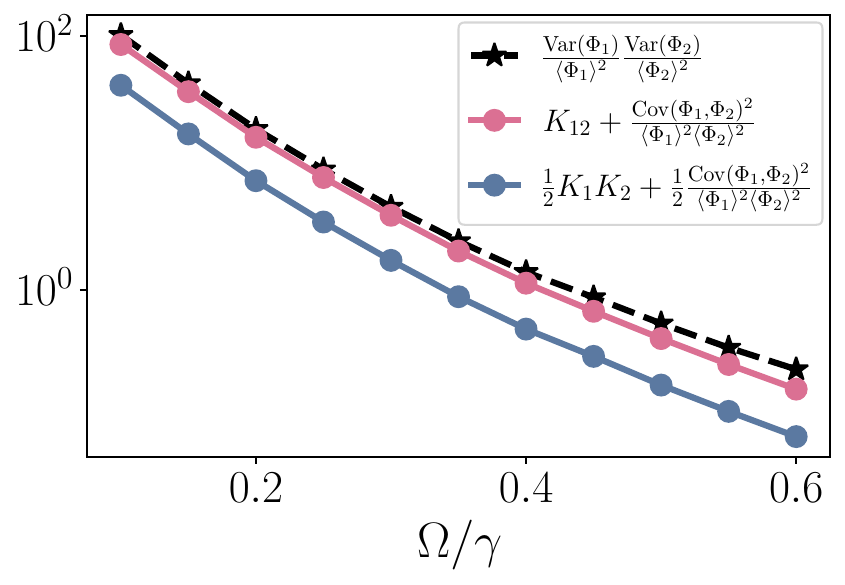}%
}
\caption{(a) Plots of the product of the relative fluctuation of the heat currents $\Theta_1$ and $\Theta_2$, and its bounds obtained from Eq.~\eqref{multidimensionalKUR}, which includes the term $K_{12}$, and from Eq.~\eqref{multiKURVu}, which includes $K_1 K_2$,
as a function of $\Omega/\gamma$ for the three-level maser. (b) Plots of the product of the relative fluctuation of the counting observables $\Phi_1$ and $\Phi_2$, the multidimensional quantum KUR bound $K_{12}$ in Eq.~\eqref{multidimensionalKUR}, and the product of quantum KUR bounds in Eq.~\eqref{quantumKUR}, $K_1 K_2$, as a function of $\Omega/\gamma$ for the three-level maser. All plots were obtained from the numerical simulation of $5 \times 10^4$ trajectories for $\tau = 10$. The values of the parameters in both cases are $\Delta = 0$, $n_1 = 5.0$ and $n_2 = 0.01$.}\label{Fig3}
\end{figure*}

 In Fig.~\ref{Fig1}(a), we depict the qubit system coupled to the bath, the arrows between them corresponding to two different time-integrated currents, $\Phi_1$ and $\Phi_2$. More precisely, each time-integrated current $\Phi_\alpha$ can be defined via the set of coefficients $\vec{w}_\alpha = \{w_k \}_{k=1}^K$, where $w_k \neq 0$ only if $k \in S_\alpha$. Since here the set containing all jump channels is given by $S = \{1,2\}$, we have that $S_1 = \{1\}$ and $S_2 = \{2\}$.
 In this way, we define $\Phi_1$ as given by $\vec{w}_1 = \{1,0\}$, and $\Phi_2$ by $\vec{w}_2 = \{0,1\}$. 
 We numerically simulate $5\times 10^4$ trajectories for $\tau = 10$, given the initial state $\rho(0) = \ket{0}\bra{0}$, and in Fig.~\ref{Fig2} we plot the product of the relative fluctuations of the two currents and our bound obtained from Eq.~\eqref{multidimensionalKUR}, which includes the term $K_{12}$ (and, therefore, the off-diagonal element of the Fisher information matrix, $F_{12}$) and a covariance term (See Appendix~\ref{monitoring_operator_formalism} for details on the simulation technique). We also plot the bound obtained from Eq.~\eqref{multiKURVu}, which also includes a covariance term, and the product $K_1K_2$ of the bounds for each individual counting observable $\Phi_1$ and $\Phi_2$.
 
 For small $\Omega/\gamma$, the rate of coherent excitation or de-excitation is very small, so that each incoherent absorption event is likely to be followed by an incoherent emission. The correlation between the two currents is thus very strong, leading to the saturation of the bound containing $K_{12}$ in Eq.~\eqref{multidimensionalKUR}. As the coherent drive strength is increased, incoherent absorption and emission become increasingly uncorrelated and the bound in Eq.~\eqref{multidimensionalKUR} becomes accordingly looser. 
 Nonetheless, the multidimensional quantum KUR bound, including the term $K_{12}$, is significantly tighter than the bound from Eq.~\eqref{multiKURVu}
 for small values of $\Omega/\gamma$, for which the correlation between the currents is strong. As the correlation between the time-integrated currents decrease as $\Omega/\gamma$ increases, the difference between the bounds becomes increasingly less pronounced.

\subsection{Three-level maser}

The three-level maser~\cite{Scovil1959, Geva1994} consists of a three-level system coupled to two thermal baths with thermal occupations $n_1$ and $n_2$.
The evolution of the system state $\rho(t)$ is described by the following quantum master equation,
\begin{multline}
    \dot{\rho}(t) =-i[H,\rho(t)] + \mathcal{D}[L_1]\rho(t) + \mathcal{D}[L_{1^\prime}]\rho(t) \\ + \mathcal{D}[L_2]\rho(t) + \mathcal{D}[L_{2^\prime}]\rho(t),
\end{multline}
where $H = \sum_{l=1}^3 \epsilon_l \ket{\epsilon_l}\bra{\epsilon_l} + \Omega (e^{i\omega_d t}\sigma_{12} + e^{-i\omega_d t}\sigma_{21})$, with $\sigma_{ij}\equiv | \epsilon_i\rangle\langle \epsilon_j|$.
The jump operators and the reverse jump operators are given by $L_1 = \sqrt{\gamma_1 n_1}\sigma_{31}$ and $L_{1^\prime} = \sqrt{\gamma_1 (1+n_1)}\sigma_{13}$, and $L_2 = \sqrt{\gamma_2 n_2}\sigma_{32}$ and $L_{2^\prime} = \sqrt{\gamma_2 (1+n_2)}\sigma_{31}$.
The strength of the couplings with the baths are given by $\gamma_1>0$ and $\gamma_2>0$, with $\Omega$ representing the strength of the external driving.
In an appropriate rotating frame, the Hamiltonian can be written as~\cite{Kalaee2021},
\begin{equation}
    \tilde{H} = \Delta\sigma_{22} + \Omega (\sigma_{12} + \sigma_{21}),
\end{equation}
where $\Delta = \omega_d - (\epsilon_2 - \epsilon_1)$ is the detuning parameter.

First, we consider time-integrated heat currents. In Fig.~\ref{Fig1}(b), we represent the two currents, $\Theta_1$ and $\Theta_2$, using arrows in different colours.
More precisely, given $S = \{1, 1^\prime, 2, 2^\prime \}$, we have that $S_1 = \{1, 1^\prime\}$ and $S_2 = \{2, 2^\prime\}$, and the set coefficients defining $\Theta_1$ are given by $\vec{w}_1 = \{1,\ -1,\ 0,\ 0\}$ and $\Theta_2$ is defined via $\vec{w}_2 = \{0,\ 0,\ 1,\ -1\}$. In Fig.~\ref{Fig3}(a), we plot the product of the relative fluctuations of the two time-integrated currents and its bound from Eq.~\eqref{multidimensionalKUR},
obtained from the numerical simulation of $5\times 10^4$ quantum trajectories for the initial state $\rho(0) = \ket{\epsilon_2}\bra{\epsilon_2}$ and $\tau = 10$. We see that our multidimensional quantum KUR is, essentially, saturated: this happens because the two currents are close to perfectly correlated for all $\Omega$. Indeed, it has been shown in Ref.~\cite{Kalaee2021} that both currents are equal, up to a constant proportionality factor, to the same stochastic variable, which represents the number of successful ``engine cycles'' of the three-level maser~\cite{Hegde2024}. This becomes even clearer by comparing the bound in Eq.~\eqref{multidimensionalKUR} with the one in Eq.~\eqref{multiKURVu}, which is plotted in blue in Fig.~\ref{Fig3}. While both bounds capture correlations, the multidimensional KUR bound in Eq.~\eqref{multiKURVu}, which is obtained from the product of the bounds  $K_1 K_2$ for each individual time-integrated current $\Theta_1$ and $\Theta_2$ in Eq.~\eqref{quantumKUR}, is significantly looser than our multidimensional bound in Eq.~\eqref{multidimensionalKUR}. 

We also consider the counting observables $\Phi_1$ and $\Phi_2$ given by $\vec{w}_1 = \{1,\ 1,\ 0,\ 0\}$ and $\vec{w}_2 = \{0,\ 0,\ 1,\ 1\}$. To obtain the plots of the relative fluctuations of both counting observables and the bounds in Eq.~\eqref{multidimensionalKUR} and Eq.~\eqref{multiKURVu}
as a function of $\Omega/\gamma$ in Fig.~\ref{Fig3}(b),  we also simulated $5 \times 10^4$ trajectories for $\tau = 10$. We see that our multidimensional bound in Eq.~\eqref{multidimensionalKUR} is not saturated in this case, and this is due to the fact that the correlations between $\Phi_1$ and $\Phi_2$ are not as strong as for the time-integrated heat currents. Furthermore, we see that it tends to become slightly looser as $\Omega$ is increased. This is due to the fact that the correlation between the observables also (slowly) decrease as $\Omega$ is increased. Finally, we see that the bound in Eq.~\eqref{multiKURVu} is much looser than the one in Eq.~\eqref{multidimensionalKUR}.




\section{Conclusion}\label{Conclusion}

In this article, we have derived new thermodynamic precision bounds, namely multidimensional KURs and TURs, encompassing multiple counting observables or currents in open quantum systems undergoing Markovian dynamics. Specifically, our new bounds incorporate contributions associated with the system's quantum evolution, as well as correlations between any two counting observables, through a covariance term. Our derivations exploit a multiparameter metrology approach, by considering a perturbed system dynamics where two different parameters are imprinted on the system's evolution. Intriguingly, the off-diagonal matrix element of the corresponding Fisher information matrix appears as a novel ingredient in this approach. This quantity vanishes for the analogous classical problem, and therefore can be considered a genuinely quantum contribution to our bounds on current-current correlations. While we considered only two counting observables here, the case of three or more observables can also be tackled straightforwardly using the same approach, albeit at the cost of more complicated expressions.

By considering a coherently-driven qubit system and a three-level maser, we performed numerical simulations and compared the genuinely multiparameter KUR from Eq.~\eqref{multidimensionalKUR} with the other multidimensional KUR bound in Eq.~\eqref{multiKURVu}, the latter having been derived from the quantum KUR for single counting observables in Eq.~\eqref{quantumKUR}. 
Although both multidimensional KURs include contributions from coherent dynamics and the covariance term, only the bound obtained from Eq.~\eqref{multidimensionalKUR} captures correlations within the off-diagonal term of the Fisher information. Therefore, as illustrated by our examples, the bound on the product of the relative fluctuations of time-integrated currents from Eq.~\eqref{multidimensionalKUR} is  generally tighter than that from Eq.~\eqref{multiKURVu}. Similarly, the genuinely multiparameter TUR~\eqref{multidimensionalTUR} is generally tighter than the analogous single-parameter TUR~\eqref{multiTURVu}.



In future work, we plan to further investigate the off-diagonal elements of the Fisher information matrix and its connections with non-classical aspects of the system dynamics, such as quantum temporal correlations and quantum invasiveness~\cite{LeggettGarg1985, Kofler2013, Knee2016, Moreira2017, Moreira2019}. 
It is known that quantum invasiveness leaves signatures in the statistics of continuous measurements~\cite{Ruskov2006,Bednorz2012}, which become even stronger when considering multiple non-commuting observables~\cite{GarciaPintos2016}. Understanding how these signatures impact current fluctuations could help to shed light on the physical origin of quantum TUR and KUR violations. In this sense, extending our results to other trajectory unravellings via the quantum Cram\'er-Rao bound~\cite{Gammelmark2014} could also prove insightful.


\section*{ACKNOWLEDGEMENTS}

We thank P. P. Potts for insightful discussions. 
 M.T.M. is supported by a Royal Society-Research Ireland University Research Fellowship. 
  M.R. acknowledges funding from Research Ireland (formerly Irish Research Council) under the Government of Ireland Postgraduate Scholarship GOIPG/2022/2321.
  F.C.B. acknowledges funding from Research Ireland (formerly Irish Research Council) (grant IRCLA/2022/3922) and the John Templeton Foundation (grant 62423).
This project is co-funded by the European Union (Quantum Flagship project ASPECTS, Grant Agreement No.~101080167) and UK Research \& Innovation (UKRI). Views and opinions expressed are however those of the authors only and do not necessarily reflect those of the European Union, Research Executive Agency or UKRI. Neither the European Union nor UKRI can be held responsible for them.
\\
\appendix

\section{Derivation of the multidimensional KUR}\label{DerivationMKUR}

To derive the multidimensional KUR bound given in Eq.~\eqref{multidimensionalKUR} from Eq.~\eqref{scalarCRbound}, we consider the following parameter imprinting on the Hamiltonian and jump operators,
\begin{align}
    & H_{\vec{\phi}} = (1 + \phi_1 + \phi_2)H, \label{phihamiltonian} \\
   &  L_{k,\vec{\phi}} = \begin{cases}
        \sqrt{1+\phi_1} L_{k} & k\in S_1 \\
        \sqrt{1+\phi_2} L_{k} & k\in S_2 
    \end{cases}. \label{phijumpop}
\end{align}
In this case, the probability distribution for observing the parameter-dependent trajectories $\gamma_{\vec{\phi}}$ are given by

\begin{widetext}
\begin{equation}
p(\gamma_{\vec{\phi}})=p_0(1+\phi_1)^{N_1}(1+\phi_2)^{N_2}|U_{\phi_1,\phi_2}(\tau,t_N)\prod_{ j=1}^{N}L_{k_{j}}U_{\phi_1,\phi_2}(t_j,t_{j-1})| \psi_0 \rangle|^2.
\end{equation}
\end{widetext}
where $N_\alpha$ corresponds to the random number of jumps in the channels $k \in S_\alpha$. In the following, we define 
\begin{equation}
\label{q_definition}
    q_{\vec{\phi}} \equiv |U_{\phi_1,\phi_2}(\tau,t_N)\prod_{j=1}^NL_{k_{j}}U_{\phi_1,\phi_2}(t_j,t_{j-1})| \psi_0 \rangle|^2.
\end{equation}
 The diagonal elements of the Fisher information matrix for $\phi_1, \phi_2$ = 0 matrix are, therefore, given by
 \begin{multline}
     F_{\alpha\alpha} = -\langle \partial_{\phi_\alpha}^2\ln(1+\phi_\alpha)^{N_\alpha} \rangle\Big|_{\vec{\phi}=0} - \langle \partial_{\phi_\alpha}^2 \ln q_{\vec{\phi}} \rangle\Big|_{\vec{\phi}=0} \\ = \langle N_\alpha\rangle + \mathcal{Q}_\alpha  \equiv \mathcal{A}_\alpha + \mathcal{Q}_\alpha, \\
 \end{multline}
 where $\mathcal{Q}_\alpha = - \langle \partial_{\phi_\alpha}^2 \ln q_{\vec{\phi}} \rangle|_{\vec{\phi}=0}$. Additionally,
the off-diagonal term of the Fisher information matrix, for $\phi_1, \phi_2$ = 0, reads
\begin{equation}
         F_{12} = \langle N_1 N_2 \rangle + \langle \partial_{\phi_1}\ln q_{\vec{\phi}} \partial_{\phi_2}\ln q_{\vec{\phi}} \rangle\Big|_{\vec{\phi}=0}.
\end{equation}

The deformed system dynamics, considering the Hamiltonian and jump operators in Eqs.~\eqref{phihamiltonian} and~\eqref{phijumpop}, reads
\begin{multline}\label{deformedME}
     \dot{\rho}_{\phi_1, \phi_2}(t) = -i[H_{\vec{\phi}},\rho_{\phi_1, \phi_2}(t)] + \sum_{k}\mathcal{D}[L_{k,\vec{\phi}}]\rho_{\phi_1, \phi_2}(t).
\end{multline}
To calculate the derivatives $\partial_{\phi_\alpha}\langle\Phi_\alpha\rangle$, we first expand the density matrix for small perturbations $\phi_1, \phi_2 \ll 1$, 
\begin{equation}\label{expandedME}
\rho_{\phi_1,\phi_2}(t) \approx \rho(t) + \zeta_1\phi_1 + \zeta_2\phi_2,
\end{equation}
where $\zeta_1$ and $\zeta_2$ are both traceless.
Replacing Eq.~\eqref{expandedME} in Eq.~\eqref{deformedME}, we get 
\begin{widetext}
\begin{multline}
    \dot{\rho}_t + \dot{\zeta_1}\phi_1 + \dot{\zeta_2}\phi_2 = -i[(1+\phi_1+\phi_2)H, \rho(t)+\zeta_1\phi_1+\zeta_2\phi_2]+\sum_{k \in S_1}(1+\phi_1)[L_{k}(\rho(t)+\phi_1\zeta_1+\phi_2\zeta_2)L_{k}^\dagger - \frac{1}{2}\{L_{k}^\dagger L_{k}, \rho(t)+\phi_1\zeta_1 + \phi_2\zeta_2\}] \\ +\sum_{k \in S_2}(1+\phi_2)[L_{k}(\rho(t)+\phi_1\zeta_1+\phi_2\zeta_2)L_{k}^\dagger - \frac{1}{2}\{L_{k}^\dagger L_{k}, \rho(t)+\phi_1\zeta_1 + \phi_2\zeta_2\}] + \mathcal{O}(\phi_1^2) + \mathcal{O}(\phi_2^2).
\end{multline}
Collecting the first order terms in $\phi_1$ and $\phi_2$, we get that $\zeta_1$ and $\zeta_2$ can be obtained via the following equations,
\begin{align}
    \dot{\zeta_1} = -i[H,\rho(t) + \zeta_1] + \sum_{k \in S_1}(\mathcal{D}[L_{k}]\rho(t) + \mathcal{D}[L_{k}]\zeta_1) + \sum_{k \in S_2}\mathcal{D}[L_{k}]\zeta_1 = \mathcal{L}_1(\rho(t)) + \mathcal{L}(\zeta_1), \label{diffzeta1}
    \\ \dot{\zeta_2} = -i[H,\rho(t) + \zeta_2] + \sum_{k \in S_2}(\mathcal{D}[L_{k}]\rho(t) + \mathcal{D}[L_{k}]\zeta_2) + \sum_{k \in S_1}\mathcal{D}[L_{k}]\zeta_2 = \mathcal{L}_2(\rho(t)) + \mathcal{L}(\zeta_2), \label{diffzeta2}
\end{align}   
where $\mathcal{L}_\alpha(\rho) \equiv -i[H, \rho] + \sum_{k \in S_\alpha} \mathcal{D}[L_{k}]\rho$.

Now using Eq.~\eqref{expandedME}, we obtain
\begin{multline}
\partial_{\phi_1}\langle \Phi_1 \rangle|_{\phi_1 = 0} \approx \partial_{\phi_1} \int_0^\tau \sum_{k \in S_1} w_{k} (1+\phi_1){\rm tr}\{L_{k}( \rho(t) + \phi_1\zeta_1 + \phi_2\zeta_2)L_{k}^\dagger\}dt|_{\phi_1  = 0} \\ = \int_0^\tau \sum_{k 
\in S_1} w_{k} {\rm tr}\{L_{k} \rho(t) L_{k}^\dagger\}dt + \int_0^\tau \sum_{k \in S_1} w_{k} {\rm tr}\{L_{k} \zeta_1 L_{k}^\dagger\}dt  = \langle \Phi_1\rangle + \langle \Phi_1^\star\rangle = (1 + \varphi_1) \langle \Phi_1\rangle,
\end{multline}
\end{widetext}
where $\varphi_1 \equiv \langle \Phi_1^\star\rangle/{\langle \Phi_1\rangle}$.
We now introduce the jump superoperators $\mathcal{J}_\alpha$, whose actuation on a operator $X$ being given by
\begin{equation}\label{superoperator}
  \mathcal{J}_\alpha X = \sum_{k \in S_\alpha} w_{k} L_{k} X L_{k}^\dagger. 
\end{equation}
In this way, we can write
\begin{equation}\label{Phialphastar}
    \langle \Phi_\alpha^\star \rangle \equiv \int_0^\tau \sum_{k \in S_\alpha} w_{k} {\rm tr}\{L_{k} \zeta_\alpha L_{k}^\dagger\}dt = \int_0^\tau dt \ {\rm tr}\{\mathcal{J}_\alpha\zeta_\alpha (t) \}.
\end{equation}

The solution to Eqs.~\eqref{diffzeta1} and~\eqref{diffzeta2} can be written as
\begin{equation}\label{zetaalpha}
    \zeta_\alpha(t) = e^{\mathcal{L}t}[\zeta_\alpha(0)] + \int dt^\prime e^{\mathcal{L}(t-t^\prime)}\mathcal{L}_\alpha e^{\mathcal{L}t^\prime}\rho(0).
\end{equation}
Thus, considering the initial condition $\zeta_\alpha(0) = 0$ and replacing $\zeta_\alpha(t)$ given in Eq.~\eqref{zetaalpha} in Eq.~\eqref{Phialphastar}, we get
\begin{multline}\label{phistar}
    \langle \Phi_\alpha^\star \rangle = \int_0^\tau dt \int_0^t dt^\prime {\rm tr}\{\mathcal{J}_\alpha e^{\mathcal{L}(t - t^\prime)}\mathcal{L}_\alpha \rho(t^\prime) \} \\ = \int_0^\tau dt \int_0^t dt^\prime \langle \mathbb{1} | \mathcal{J}_\alpha e^{\mathcal{L}(t-t^\prime)} \mathcal{L}_\alpha e^{\mathcal{L}t^\prime}|\rho(0)   \rangle.
\end{multline}

We note that we switched to the vectorization notation in the equation above, where $\ket{\rho}$ is the vectorized density matrix and $\ket{\mathbb{1}}$ satisfies ${\rm tr}\{ \rho \} = \langle \mathbb{1} | \rho \rangle$~\cite{Landi2024}. 
In the long time limit, we can write Eq.~\eqref{phistarvec} as
\begin{widetext}
\begin{multline}\label{phistarvec}
    \langle \Phi_\alpha^\star \rangle = \int_0^\tau dt \int_0^t dt^\prime \langle \mathbb{1} | \mathcal{J}_\alpha \left[\ket{\rho_{\text{ss}}}\bra{\mathbb{1}} + \sum_{j\neq 0} e^{\lambda_j(t-t^\prime)}\ket{x_j}\bra{y_j}\right] \mathcal{L}_\alpha \left[\ket{\rho_{\text{ss}}}\bra{\mathbb{1}} + \sum_{k\neq 0} e^{\lambda_k t^\prime} \ket{x_k}\bra{y_k}\right] |\rho(0)\rangle \\
    =  \int_0^\tau dt \int_0^t dt^\prime \left[ \sum_{j\neq 0} e^{\lambda_j(t-t^\prime)} \langle \mathbb{1}| \mathcal{J}_\alpha | x_j  \rangle \langle y_j| \mathcal{L}_\alpha|\rho_{\text{ss}} \rangle + \sum_{j,k\neq 0} e^{\lambda_jt} e^{(\lambda_k - \lambda_j) t^\prime} \langle \mathbb{1}| \mathcal{J}_\alpha | x_j \rangle \langle y_j| \mathcal{L}_\alpha |x_k \rangle \langle y_k | \rho(0) \rangle \right],
\end{multline}  
where we expressed the matrix exponential appearing in Eq.~\eqref{phistar} as $e^{\mathcal{L} t} = \ket{\rho_{\text{ss}}}\bra{\mathbb{1}} + \sum_{j\neq 0} e^{\lambda_j t}\ket{x_j}\bra{y_j}$, with $\lambda_j$ being eigenvalues of the Liouvillian $\mathcal{L}$, with correspondent right and left eigenvectors given by $\ket{x_j}$ and $\bra{y_j}$, respectively~\cite{Landi2024}.
We can further evaluate Eq.~\eqref{phistarvec} by noting that the integration of the second term gives
\begin{multline}\label{firstterm}
    \int_0^\tau dt \int_0^t dt^\prime  \sum_{j,k\neq 0} e^{\lambda_jt} e^{(\lambda_k - \lambda_j) t^\prime} \langle \mathbb{1}| \mathcal{J}_\alpha | x_j \rangle \langle y_j| \mathcal{L}_\alpha |x_k \rangle \langle y_k | \rho(0) \rangle  = -
   \sum_{j, k\neq 0}\langle \mathbb{1}| \mathcal{J}_\alpha \frac{1}{\lambda_j}| x_j \rangle \langle y_j| \mathcal{L}_\alpha \frac{1}{\lambda_k} |x_k \rangle \langle y_k | \rho(0) \rangle \\ = - \langle \mathbb{1}|\mathcal{J}_\alpha \mathcal{L}^+\mathcal{L}_\alpha\mathcal{L}^+ |\rho(0) \rangle = O(\tau^0),
\end{multline}
where $\mathcal{L}^+$ is the {\it Drazin inverse} of $\mathcal{L}$, given by
\begin{equation}\label{Drazin}
    \mathcal{L}^+ = \sum_{j\neq 0} \frac{1}{\lambda_j} \ket{x_j}\bra{y_j} = -\int_0^\infty dt \ e^{\mathcal{L}t}(1 - \ket{\rho_{\text{ss}}}\bra{\mathbb{1}}).
\end{equation}
By integrating the first term of Eq.~\eqref{phistarvec}, we obtain
\begin{equation}
    \int_0^\tau dt \int_0^t dt^\prime  \sum_{j\neq 0} e^{\lambda_j(t-t^\prime)} \langle \mathbb{1}| \mathcal{J}_\alpha | x_j  \rangle \langle y_j| \mathcal{L}_\alpha|\rho_{\text{ss}} \rangle \approx -\tau \sum_{j\neq 0} \frac{1}{\lambda_j} \langle \mathbb{1} | \mathcal{J}_\alpha | x_j \rangle \langle y_j | \mathcal{L}_\alpha | \rho \rangle = -\tau \langle \mathbb{1} | \mathcal{J}_\alpha \mathcal{L}^+\mathcal{L}_\alpha | \rho_{\text{ss}} \rangle.
\end{equation}
\end{widetext}
As a result,
\begin{equation}\label{phistardrazin}
    \langle \Phi^\star_\alpha \rangle \approx -\tau \langle \mathbb{1} | \mathcal{J}_\alpha \mathcal{L}^+\mathcal{L}_\alpha | \rho_{\text{ss}} \rangle.
\end{equation}
Furthermore, using the expression for the Drazin inverse from Eq.~\eqref{Drazin} in Eq.~\eqref{phistardrazin}, we can rewrite it as
\begin{widetext}
\begin{multline}\label{Phistarsolution}
    \langle \Phi^\star_1 \rangle = \tau \int_0^\infty dt \langle  \mathbb{1} | \mathcal{J}_1 (e^{\mathcal{L}t} - e^{\mathcal{L}t}\ket{\rho_{\text{ss}}}\bra{\mathbb{1}}) \mathcal{L}_1 | \rho_{\text{ss}}  \rangle = \tau\int_0^\infty dt \ {\rm tr}\{ \mathcal{J}_1 e^{\mathcal{L}t} \mathcal{L}_1 \rho_{\text{ss}} \} 
    = -\tau\int_0^\infty dt \ {\rm tr}\{ \mathcal{J}_1 e^{\mathcal{L}t} \mathcal{D}_2 \rho_{\text{ss}}\},
\end{multline}
where we used that ${\rm tr}\{ \mathcal{L}_\alpha \rho_{{\text{ss}}} \} = 0$, and $\mathcal{D}_2\rho_\text{ss} = \mathcal{L}\rho_\text{ss} - \mathcal{L}_1\rho_\text{ss} = - \mathcal{L}_1\rho_\text{ss}$ and $\mathcal{D}_2\rho = \sum_{k \in S_2}\mathcal{D}[L_{k}]\rho$.
Now, we can write
\begin{multline}
    {\rm tr}\{ \mathcal{J}_1 e^{\mathcal{L}t} \mathcal{D}_2 \rho_{\text{ss}}\} = \sum_{k \in S_1} \sum_{\bar{k}\in S_2} w_{k} {\rm tr} \{e^{\mathcal{L}^+t} L_{k}^\dagger L_{k} (L_{\bar{k}}\rho_{\text{ss}}L_{\bar{k}}^\dagger -\frac{1}{2} L_{\bar{k}}^\dagger L_{\bar{k}} \rho_{\text{ss}} - \frac{1}{2} \rho_{\text{ss}} L_{\bar{k}}^\dagger  L_{\bar{k}}) \} \\ = \sum_{k \in S_2} \left\langle L_{k}^\dagger W_1(t)L_{k} - \frac{1}{2}W_1(t) L_{k}^\dagger L_{k} - \frac{1}{2} L_{k}^\dagger  L_{k} W_1(t) \right\rangle_{\text{ss}} = \frac{1}{2} \sum_{k \in S_2} \left\langle  [L_{k}^\dagger, W_1(t)] L_{k} + L_{k}^\dagger [W_1(t), L_{k}] \right\rangle_{\text{ss}},
\end{multline}
where we defined $W_1 \equiv \sum_{k \in S_1} L_{k}^\dagger L_{k} w_{k}$ and $\langle X \rangle_{\text{ss}} \equiv {\rm tr}\{ X \rho_{\text{ss}}\}$. As a result, we finally obtain
\begin{equation}
    \langle \Phi^\star_1 \rangle \approx - \frac{\tau}{2} \sum_{k \in S_2} \int_0^\infty dt \left\langle  [L_{k}^\dagger, W_1(t)] L_{k} + L_{k}^\dagger [W_1(t), L_{k}] \right\rangle_{\text{ss}}.
\end{equation}

Thus, from Eq.\eqref{scalarCRbound}, we get the following multidimensional KUR,
\begin{equation}
      \frac{{\rm Var}(\Phi_1)}{\langle \Phi_1\rangle^2} \frac{{\rm Var(\Phi_2)}}{\langle\Phi_2\rangle^2} \geq  \frac{(1+\varphi_1)^2(1+\varphi_2)^2}{(\mathcal{A}_1+\mathcal{Q}_1)(\mathcal{A}_2+\mathcal{Q}_2)-F_{12}^2} + \frac{\rm Cov(\Phi_1,\Phi_2)^2}{\langle \Phi_1\rangle^2\langle\Phi_2\rangle^2}.
\end{equation}
\end{widetext}

\section{Derivation of the multidimensional TUR}\label{DerivationMTUR}

\begin{widetext}
We now derive the multidimensional TUR
in Eq.~\eqref{multidimensionalTUR}. First, we show that the entropy production can be expressed as

\begin{align}
\dot{\Sigma} & = -\text{tr}\{\dot{\rho}(t)\ln\rho(t)\} + \sum_{k \in  S_1} \text{tr}\{L_{k}^\dagger L_{k}\}\Delta s_{k} + \sum_{k \in S_2} \text{tr}\{L_{k}^\dagger L_{k}\}\Delta s_{k} \notag \\
& =
-\sum_{k \in S_1} \text{tr}\{\mathcal{D}[L_{k}]\rho(t)\ln\rho(t)\} -\sum_{k \in S_2} \text{tr}\{\mathcal{D}[L_{k}]\rho(t)\ln\rho(t)\} +\sum_{k \in S_1} \text{tr}\{L_{k}^\dagger L_{k}\}\Delta s_{k} + \sum_{k \in S_2} \text{tr}\{L_{k}^\dagger L_{k}\}\Delta s_{k} \notag  \\ & = \sum_{k \in S_1}{\rm tr}\{L_{k}\rho(t)(\Delta s_{k}L_{k}^\dagger - [L_{k}^\dagger,\ln\rho(t)]) \} + \sum_{k \in S_2}{\rm tr}\{L_{k}\rho(t)(\Delta s_{k}L_{k}^\dagger - [L_{k}^\dagger,\ln\rho(t)]) \} = \dot{\Sigma}_1 + \dot{\Sigma}_2,
\end{align}
\end{widetext}

where $\Delta s_{k} = -\Delta s_{k^\prime}$, and
\begin{equation}
    \Sigma_\alpha \equiv \int_0^\tau dt \sum_{k\in S_\alpha}{\rm tr}\{L_{k}\rho(t)(\Delta s_{k}L_{k}^\dagger - [L_{k}^\dagger,\ln\rho(t)]) \}.
\end{equation}

The derivation of the multidimensional TUR in Eq.~\eqref{multidimensionalTUR} also follows from Eq.~\eqref{scalarCRbound} for the following encoding of the parameters $\theta_1$ and $\theta_2$, 
\begin{align}
    & H_{\vec{\theta}} = (1 + \theta_1 + \theta_2)H, \label{thetahamiltonian} \\
    & L_{k,\vec{\theta}} =\begin{cases}
         \sqrt{1+ l_{k}(t)\theta_1} L_{k} & k\in S_1 \\
        \sqrt{1+ l_{k}(t)\theta_2} L_{k} & k\in S_2 
    \end{cases}, 
\end{align}
As local detailed balance is assumed, we note that all pairs of channels $k, k^\prime$ satisfying $L_k = e^{\Delta s_k/2}L_{k^\prime}^\dagger$, will belong to the same set $S_\alpha$.
The probability distribution of observing the trajectories $\gamma_{\vec{\theta}}$ in the deformed dynamics is given by
\begin{widetext}
\begin{equation}
       p(\gamma_{\vec{\theta}})=p_0 \prod_{j \in T_1}^{} [1+ l_{k_j} (t_{j})\theta_1]\prod_{j \in T_2}^{}[1+ l_{k_j}(t_{j}) \theta_2]|U_{\theta_1,\theta_2}(\tau,t_N)\prod_{j}^NL_{k_{j}}U_{\theta_1, \theta_2}(t_j,t_{j-1})| \psi_0 \rangle|^2,
\end{equation}
where $T_\alpha$ denotes the set of jumps $j$ occurring in channels $k$ belonging to $S_\alpha$, i.e. $j \in T_\alpha$ if $k \in S_\alpha$.
\begin{equation}\label{lkalpha}
    l_{k}(t) = \frac{{\rm tr}\{L_{k}\rho(t) L_{k}^\dagger\}- {\rm tr}\{L_{k^\prime}\rho(t)L_{k^\prime}^\dagger\}}{{\rm tr}\{L_{k}\rho(t)L_{k}^\dagger\} + {\rm tr}\{L_{k^\prime}\rho(t)L_{k^\prime}^\dagger\}}.
\end{equation}
The elements of the Fisher information matrix can therefore be obtained as follows,
    \begin{multline}
    F_{\alpha \alpha} = -\left\langle \partial_{\theta_\alpha}^2\ln\prod_{j \in T_\alpha} [1+ l_{k_j} (t_{j})\theta_\alpha]\right\rangle \Bigg|_{\vec{\theta} = \vec{0}} - \langle \partial_{\theta_\alpha}^2 \ln q_{\theta_1, \theta_2} \rangle\Big|_{\vec{\theta}= \vec{0}} = \left\langle \sum_{j \in T_\alpha} {l_{k_j}^2(t_j)} \right\rangle + \mathcal{Q}_{\alpha}^\prime \\
    = \frac{1}{2} \int_0^\tau dt \sum_{k \in S_\alpha} [{\rm tr}\{ L_{k} \rho(t) L_{k}^\dagger \} + {\rm tr}\{ L_{k^\prime} \rho(t) L_{k^\prime} \}] l_{k}^2(t) + \mathcal{Q}_\alpha^\prime = \frac{1}{2} \int_0^\tau dt \sum_{k \in S_\alpha} \frac{[{\rm tr}\{ L_{k} \rho(t) L_{k}^\dagger \} - {\rm tr}\{ L_{k^\prime} \rho(t) L_{k^\prime}^\dagger\}]^2 }{  {\rm tr}\{L_{k} \rho(t) L_{k}^\dagger \} + {\rm tr}\{ L_{k^\prime} \rho(t) L_{k^\prime}^\dagger\}}  + \mathcal{Q}^\prime.
\end{multline}
We note that, since from Ref.~\cite{VanVu2022},
\begin{equation}
    \dot{\Sigma} = \dot{\Sigma}_1 + \dot{\Sigma}_2 \geq \sum_{k \in S_1}\dfrac{ [{\rm tr}\{L_{k}\rho(t) L_{k}^\dagger\} - {\rm tr}\{L_{k^\prime}\rho(t) L_{k^\prime}^\dagger\}]^2}{{\rm tr}\{L_{k}\rho(t) L_{k}^\dagger\} + {\rm tr}\{L_{k^\prime}\rho(t) L_{k^\prime}^\dagger\}} + \sum_{k \in S_2}\dfrac{ [{\rm tr}\{L_{k}\rho(t) L_{k}^\dagger\} - {\rm tr}\{L_{k^\prime}\rho(t) L_{k^\prime}^\dagger\}]^2}{{\rm tr}\{L_{k}\rho(t) L_{k}^\dagger\} + {\rm tr}\{L_{k^\prime}\rho(t) L_{k^\prime}^\dagger\}},
\end{equation}
\end{widetext}
we can conclude that
\begin{equation}
    F_{\alpha \alpha} \leq \Sigma_\alpha/2 + \mathcal{Q}_\alpha^\prime.
\end{equation}
In turn, the off-diagonal element of the Fisher information matrix is given by
\begin{multline}
    F_{12} = \left\langle \sum_{j \in T_1}^{} l_{k_j}(t_j) \sum_{j \in T_2}^{}l_{k_j}(t_j) \right\rangle \\ + \langle \partial_{\theta_1}\ln q_{\vec{\theta}} \partial_{\theta_2}\ln q_{\vec{\theta}} \rangle\Big|_{\vec{\theta}=\vec{0}},
\end{multline}
where $q_{\theta_1, \theta_2} \equiv |U_{\theta_1, \theta_2}(\tau,t_N)\prod_{j}^NL_{k_j}U_{\theta_1, \theta_2}(t_j,t_{j-1})| \psi_0 \rangle|^2$.
Note that the first term in the expression above can be expressed as
\begin{multline}
    \left\langle \sum_{j \in T_1}^{} l_{k_j}(t_{j}) \sum_{j \in T_2}^{}l_{k_j}(t_{j}) \right\rangle \\ = \int_0^\tau dt_1 \sum_{k \in S_1} l_{k}(t_1) \int_0^\tau dt_2 \sum_{\bar{k} \in S_2} l_{\bar{k}}(t_2) \langle I_{k}(t_1) I_{\bar{k}}(t_2) \rangle,
\end{multline}
where $I_{k}(t)\equiv \frac{dN_k(t)}{dt}$, $N_k$ being the number of jumps in channel $k$.

The system dynamics deformed by the parameters $\theta_1$ and $\theta_2$ will read
\begin{equation}\label{thetadeformedME}
     \dot{\rho}_{\theta_1, \theta_2}(t) = -i[H_{\vec{\theta}},\rho_{\theta_1, \theta_2}(t)] + \sum_{k}\mathcal{D}[L_{k, \vec{\theta}}]\rho_{\theta_1, \theta_2}(t). 
\end{equation}
  As in Appendix~\ref{DerivationMKUR}, to calculate the derivatives $\partial_{\theta_\alpha}\langle\Theta_\alpha\rangle$, we first expand the deformed density matrix for small perturbations $\theta_1, \theta_2 \ll 1$, 
\begin{equation}\label{thetaexpandedME}
\rho_{\phi_1,\phi_2}(t) \approx \rho(t) + \xi _1\theta_1 + \xi_2\theta_2,
\end{equation}
where $\xi_1$ and $\xi_2$ are both traceless.
Replacing this in Eq.~\eqref{thetadeformedME}, we get 
\begin{widetext}
\begin{multline}
    \dot{\rho}_t + \dot{\xi_1}\theta_1 + \dot{\xi_2}\theta_2 = -i[(1+\theta_1+\theta_2)H, \rho(t)+\theta_1+\theta_2]+\sum_{k \in S_1}(1+ l_{k}(t)\theta_1)[L_{k}(\rho(t)+\theta_1\zeta_1+\theta_2\zeta_2)L_{k}^\dagger - \frac{1}{2}\{L_{k}^\dagger L_{k}, \rho(t)+\phi_1\xi_1 + \theta_2\xi_2\}] \\ +\sum_{k \in S_2}(1+ l_{k}(t)\theta_2)[L_{k}(\rho(t)+\theta_1\xi_1+\theta_2\xi_2)L_{k}^\dagger - \frac{1}{2}\{L_{k}^\dagger L_{k}, \rho(t)+\theta_1\xi_1 + \theta_2\xi_2\}] + \mathcal{O}(\theta_1^2) + \mathcal{O}(\theta_2^2).
\end{multline}
Collecting the first order terms in $\theta_1$ and $\theta_2$, we get that $\xi_1$ and $\xi_2$ should satisfy
\begin{align}
    \dot{\xi}_1 = -i[H,\rho(t) + \xi_1(t)] + \sum_{k \in S_1}l_{k}(t)\mathcal{D}[L_{k}]\rho(t) + \sum_{k \in S_1}\mathcal{D}[L_{k}]\xi_1(t) + \sum_{k \in S_2}\mathcal{D}[L_{k}]\xi_1(t) = \mathcal{L}_1[\rho(t)]\rho(t) 
    +\mathcal{L}\xi_1(t),
    \label{diffxi1}\\ \dot{\xi}_2 = -i[H,\rho(t) + \xi_2(t)] + \sum_{k\in S_2}l_{k}(t)\mathcal{D}[L_{k}]\rho(t) + \sum_{k\in S_2}\mathcal{D}[L_{k}]\xi_2(t) + \sum_{k\in S_1}\mathcal{D}[L_{k}]\xi_2(t) = \mathcal{L}_2[\rho(t)]\rho(t)+\mathcal{L}\xi_2(t). \label{diffxi2}
\end{align}
\end{widetext}
In the equations above, we defined $\mathcal{L}_\alpha[\rho(t)]\rho \equiv  -i[H, \rho] + \sum_{k\ \in S_\alpha} l_{k}(t)\mathcal{D}[L_{k}]\rho $, denoting the dependence of the Liouvillian with $\rho(t)$ via $l_{k}(t)$, see Eq.~\eqref{lkalpha}.
As we are dealing with current observables, defined in Eq.~\eqref{multicurrents}, we have that $w_{k} = -w_{k}$. Also, note that $w_{k} l_{k}(t) = w_{k^\prime} l_{k^\prime}(t)$. Using these relations and Eq.~\eqref{lkalpha}, we can compute the derivatives of the average currents for $\theta_1, \theta_2 = 0$ as follows
\begin{widetext}
 \begin{multline}
\partial_{\theta_1}\langle \Theta_1 \rangle|_{\theta_1 = 0} \approx \partial_{\theta_1} \int_0^\tau \sum_{k \in S_1} w_{k} (1+l_{k}(t)\theta_1){\rm tr}\{L_{k}( \rho(t) + \theta_1\xi_1 + \theta_2\xi_2)L_{k}^\dagger\}dt|_{\theta_1  = 0} \\ = \int_0^\tau \sum_{k \in S_1} w_{k} l_{k}(t) {\rm tr}\{L_{k} \rho(t) L_{k}^\dagger\}dt + \int_0^\tau \sum_{k \in S_1} w_{k} {\rm tr}\{L_{k} \xi_1 L_{k}^\dagger\}dt \\ = \int_0^\tau \sum_{k, k^\prime \in S_1}  w_{k} l_{k}(t) [ {\rm tr}\{L_{k} \rho(t) L_{k}^\dagger\} + {\rm tr}\{L_{k^\prime} \rho(t) L_{k^\prime}^\dagger\} ] dt + \int_0^\tau \sum_{k \in S_1} w_{k} {\rm tr}\{L_{k} \xi_1 L_{k}^\dagger\}dt \\ = \int_0^\tau \sum_{k, k^\prime \in S_1}  w_{k} [ {\rm tr}\{L_{k} \rho(t) L_{k}^\dagger\} - {\rm tr}\{L_{k^\prime} \rho(t) L_{k^\prime}^\dagger\} ] dt + \int_0^\tau \sum_{k \in S_1} w_{k} {\rm tr}\{L_{k} \xi_1L_{k}^\dagger\}dt \\ = \int_0^\tau  \sum_{k \in S_1} w_{k}  {\rm tr}\{L_{k} \rho(t) L_{k}^\dagger\}  dt + \int_0^\tau \sum_{k \in S_1} w_{k} {\rm tr}\{L_{k} \xi_1 L_{k}^\dagger\}dt
\\ =
\langle \Theta_1\rangle + \langle \Theta_1^\star\rangle = (1 + \vartheta_1) \langle \Theta_1\rangle,
\end{multline}
\end{widetext}
where $\vartheta_1 \equiv \langle \Theta_1 \rangle / \langle \Theta_1^\star \rangle $. Using the jump superoperator in Eq.~\eqref{superoperator}, we can write
\begin{equation}
    \langle \Theta_\alpha^\star \rangle = \int_0^\tau dt \ {\rm tr}\{\mathcal{J}_\alpha\xi_\alpha (t) \}.
\end{equation}
The solution to Eqs.~\eqref{diffxi1} and~\eqref{diffxi2} can be expressed as
\begin{equation}\label{xialpha}
    \xi_\alpha(t) = e^{\mathcal{L}t}[\xi_\alpha(0)] + \int_0^t dt^\prime e^{\mathcal{L}(t-t^\prime)}\mathcal{L}_\alpha[\rho(t')] e^{\mathcal{L}t^\prime}\rho(0).
\end{equation}
Since $\xi_\alpha(0) = 0$, the first element on the right-hand side of equation above is equal to zero.
Note that in the long time limit $\rho(t) = \rho_{\text{ss}}$ and the Liouvillian $\mathcal{L}_\alpha[\rho(t)] = \mathcal{L}_\alpha[\rho_\text{ss}]$ will be independent of time.
We now assume that the system starts at the stationary state, i.e. $\rho(0) = \rho_{\text{ss}}$. Then, Eq.~\eqref{xialpha} reads
\begin{equation}
    \xi_\alpha = \int_0^t dt^\prime e^{\mathcal{L}(t-t^\prime)}\mathcal{L}_\alpha[\rho_\text{ss}] e^{\mathcal{L}t^\prime}\rho_\text{ss}.
\end{equation}
Since the Liouvillians $\mathcal{L}_\alpha[\rho_\text{ss}]$ do not depend on time, 
we can now follow the same steps as in Appendix~\ref{DerivationMKUR}, used to derive Eq.~\eqref{Phistarsolution}, to obtain
\begin{equation}\label{thetastaraverage}
    \langle \Theta_1^\star \rangle = \int_0^\infty dt \ {\rm tr}\{ \mathcal{J}_1 e^{\mathcal{L}t} \mathcal{L}_1[\rho_\text{ss}] \rho_{\text{ss}} \}.
\end{equation}
In the following, it will be useful to define the superoperator $\mathcal{L}[\rho]\rho = -i[H, \rho] + \sum_{k} l_{k}(t)\mathcal{D}[L_{k}]\rho$. By denoting $l_{k}(t)$ in Eq.~\eqref{lkalpha} at the stationary state by $l_{k}^\text{ss}$, we can then write $\mathcal{D}_2[\rho_{\text{ss}}] \rho_\text{ss} = \mathcal{L}[\rho_{\text{ss}}]\rho_\text{ss} - \mathcal{L}_1[\rho_\text{ss}]\rho_\text{ss} = - \mathcal{L}_1[\rho_\text{ss}]\rho_\text{ss}$, and Eq.~\eqref{thetastaraverage} becomes
\begin{equation}\label{thetastarss}
    \langle \Theta_1^\star \rangle = -\int_0^\infty dt \ {\rm tr}\{ \mathcal{J}_1 e^{\mathcal{L}t} \mathcal{D}_2[\rho_\text{ss}] \rho_{\text{ss}} \}.
\end{equation}
As in Appendix~\ref{DerivationMKUR}, we can further calculate Eq.~\eqref{thetastarss} as follows
\begin{widetext}
\begin{multline}
    {\rm tr}\{ \mathcal{J}_1 e^{\mathcal{L}t} \mathcal{D}_2[\rho_\text{ss}] \rho_{\text{ss}}\} = \sum_{\bar{k} \in S_1} \sum_{k \in S_2} l_{k}^\text{ss} w_{\bar{k}} {\rm tr} \{e^{\mathcal{L}^+t} L_{\bar{k}}^\dagger L_{\bar{k}} (L_{k}\rho_{\text{ss}}L_{k}^\dagger -\frac{1}{2} L_{k}^\dagger L_{k} \rho_{\text{ss}} - \frac{1}{2} \rho_{\text{ss}} L_{k}^\dagger  L_{k}) \} \\ = \sum_{k \in S_2} l_{k}^\text{ss} \left\langle L_{k}^\dagger W_1(t)L_{k} - \frac{1}{2}W_1(t) L_{k}^\dagger L_{k} - \frac{1}{2} L_{k}^\dagger  L_{k} W_1(t) \right\rangle_{\text{ss}} = \frac{1}{2} \sum_{k\in S_2} l_{k}^\text{ss} \left\langle  [L_{k}^\dagger, W_1(t)] L_{k} + L_{k}^\dagger [W_1(t), L_{k}] \right\rangle_{\text{ss}}. 
\end{multline}
Therefore, we get
\begin{equation}
    \langle \Theta^\star_1 \rangle = - \frac{\tau}{2} \sum_{k \in S_2} l_{k}^{\text{ss}}\int_0^\infty dt \left\langle  [L_{k}^\dagger, W_1(t)] L_{k} + L_{k}^\dagger [W_1(t), L_{k}] \right\rangle_{\text{ss}}.
\end{equation}

As a result, from Eq.\eqref{scalarCRbound}, we get the following the multidimensional TUR,
\begin{equation}
      \frac{{\rm Var}(\Theta_1)}{\langle \Theta_1\rangle^2} \frac{{\rm Var(\Theta_2)}}{\langle\Theta_2\rangle^2} \geq \frac{2(1+\vartheta_1)^2(1+\vartheta_2)^2}{(\Sigma_1+2\mathcal{Q}^\prime_1)(\Sigma_2+2\mathcal{Q}^\prime_2)-2F_{12}^2}+ \frac{\rm Cov(\Theta_1, \Theta_2)^2}{\langle \Theta_1\rangle^2\langle\Theta_2\rangle^2}.
\end{equation}

\end{widetext}

\section{Multi-parameter Fisher information in classical Markovian systems}\label{Offdiagonal}

We now show that the multi-parameter Fisher information matrix for a classical master equation in the multi-bath case is diagonal. We consider a scenario where the system density matrix is diagonal at all times in a given preferred basis of orthogonal states, labelled by the Greek index $\sigma = 1,\ldots,d$. The master equation describing this scenario is given by Eq.~\eqref{ME}, with a Hamiltonian that is diagonal in the preferred basis, $H=\sum_\sigma E_\sigma \ket{\sigma}\bra{\sigma}$, and jump operators of the form $L_k = \sqrt{R_{\mu\sigma}}\ket{\mu}\bra{\sigma}$. Here, the super-index $k = \sigma\to \mu$ labels transitions between distinct pairs of states that occur with rate $R_{\mu\sigma}$ (we define $R_{\sigma\sigma} = 0$). 

Assuming that the initial state is diagonal,~i.e.~$[\rho(0),H]=0$, the solution of the master equation at any time is also of the diagonal form $\rho(t)=\sum_\sigma p_\sigma(t)\ket{\sigma}\bra{\sigma}$. The probabilities $p_\sigma(t)$ obey the classical master equation
\begin{equation}
\dot{p}_{\sigma}  = \sum_{\mu} R_{\mu\sigma}p_\mu - \Gamma_\sigma p_\sigma,
\end{equation}
where $\Gamma_\sigma = \sum_{\mu =1}^d R_{\mu\sigma}$ is the total escape rate from the state $\sigma$. 

Since the final state of jump $k_j$ is the initial state of the next jump $k_{j+1}$, a stochastic trajectory with $N$ jumps can equivalently be specified by the sequence of states and jump times, as $\gamma_{\{0,\tau\}} = \{(t_0,\sigma_0),(t_1,\sigma_1),\ldots,(t_N,\sigma_N)\}$, where $\sigma_0$ is the state at the initial time $t= t_0$, $\sigma_1$ is the state after the first jump at $t=t_1$, and so on. The probability of observing such a trajectory within a total time $t$ is given by~\cite{Prech2024}
\begin{equation}
\label{classical_path_prob}
p(\gamma_{\{0,\tau\}}) = \left(\prod_{j=0}^{N-1} R_{\sigma_{j+1}\sigma_{j}}\right) e^{-\sum_{j=0}^N\Gamma_{\sigma_{j}}(t_{j+1}-t_{j})} p_{\sigma_0},
\end{equation}
where $p_{\sigma_0}$ denotes the initial probability distribution (usually taken to be the steady state), and we define $t_{N+1} = \tau$ and $t_0 = 0$. 

We assume that each distinct transition $\mu\to\sigma$ is driven by coupling to a unique reservoir, labelled by the index $\alpha = 1,\ldots,D$, where $D$ is the number of baths. The transition rates can therefore be written as
\begin{equation}
R_{\sigma\mu} = \sum_{\alpha=1}^D R_{\sigma\mu}^{(\alpha)}, \quad\quad \Gamma^{(\alpha)}_\sigma = \sum_{\mu=1}^d R^{(\alpha)}_{\mu\sigma},
\end{equation}
where, for a given $\mu$ and $\sigma$, $R^{(\alpha)}_{\mu\sigma}$ is non-zero for only a single value of $\alpha$ corresponding to the bath driving the transition $\sigma\to\mu$. We now deform the transition rates by a different parameter for each bath, $R_{\mu\sigma}^{(\alpha)}\to (1+\phi_\alpha)R_{\mu\sigma}$. The deformed path probability $p_{\vec{\phi}}(\gamma_{\{0,\tau\}})$ is related to the undeformed probability distribution~\eqref{classical_path_prob} by
\begin{equation}
\label{classical_deformed_path_prob}
    \frac{p_{\vec{\phi}}(\gamma_{\{0,\tau\}})}{p(\gamma_{\{0,\tau\}})} = \prod_{\alpha=1}^D \left(1+\phi_\alpha\right)^{N_\alpha} e^{-\phi_\alpha \sum_{j=0}^N\Gamma_{\sigma_{j}}(t_{j+1}-t_{j})},
\end{equation}
where $N_\alpha$ is the total number of jumps induced by coupling to bath $\alpha$. More precisely, we have
\begin{equation}
    N_\alpha = \sum_{j=0}^{N-1} \frac{R^{(\alpha)}_{\sigma_{j+1}\sigma_j}}{R_{\sigma_{j+1}\sigma_j}},
\end{equation}
where, following our assumptions, each term in the sum equals one if the jump $\sigma_j\to\sigma_{j+1}$ is driven by bath $\alpha$, and is zero otherwise.

Now, using Eq.~\eqref{classical_deformed_path_prob}, it is straightforward to compute the elements of the Fisher information matrix, given by~\cite{Kay1993}
\begin{align}
    F_{\alpha\beta}(\vec{\phi}) & = -\left \langle \frac{\partial^2}{\partial\phi_\beta\partial\phi_\alpha} \ln p_{\vec{\phi}} (\gamma_{\{0,\tau\}}) \right\rangle \notag \\ 
    & = -\left \langle \frac{\partial}{\partial\phi_\beta} \left(\frac{N_\alpha}{1+\phi_\alpha} - \sum_{j=0}^N \Gamma_{\sigma_j}(t_{j+1}-t_j)\right)\right\rangle \notag \\
    & = \frac{ \langle N_\alpha\rangle}{(1+\phi_\alpha)^2}\delta_{\alpha\beta}.
\end{align}
Taking the parameters to zero, $\phi_\alpha \to 0$, we find the simple result
\begin{equation}
    F_{\alpha\beta}(\vec{\phi}=0) = \mathcal{A}_\alpha \delta_{\alpha\beta},
\end{equation}
where the dynamical activity is defined by $\mathcal{A}_\alpha = \langle N_\alpha\rangle$.

\section{Simulation technique}\label{monitoring_operator_formalism}
Inequalities such as the Cramér-Rao bound in Eq.~\eqref{CRbound} require the computation of the Fisher information contained in the measurement record of continuously monitored quantum systems. This calculation cannot be tackled directly, by using the definition of Eq.~\eqref{FI}, because the dimensionality of the summation space grows exponentially with time. To solve this problem, one can exploit the monitoring operator formalism, originally introduced in~\cite{Gammelmark2013} and recently systematised in~\cite{Radaelli2023}. We give here a quick overview of the method, which was exploited in the simulations of this work.

Let us introduce the unnormalised conditional state of the system at time $t$, $\ket{\tilde{\psi}(t)}$, given by the repeated application of the Kraus operators of Eq.~\eqref{KrausOp} to the initial state $\ket{\psi_0}$. The probability of a given measurement record is given by
\begin{equation}
    p_{\vec{\phi}}(\gamma) = ||M_{x(\tau)} M_{x(\tau - dt)}\ldots M_{x(0)}\ket{\psi_0}||^2,
\end{equation}
where $x(t)$ denotes the label of the Kraus operator to be chosen at time $t$. Let us define the  parameter specific \textit{monitoring operators} $\xi^{(\alpha)}_t$ as
\begin{equation}
    \xi^{(\alpha)}_t = \frac{\left(\partial_{\phi_\alpha} \ket{\tilde{\psi}(t)}\right)\bra{\tilde{\psi}(t)} + \ket{\tilde{\psi}(t)}\left(\partial_{\phi_\alpha}\bra{\tilde{\psi}(t)}\right)}{\left \lVert\ket{\tilde{\psi}(t)}\right\rVert^2}.
\end{equation}
It can be proven~\cite{Radaelli2023} that
\begin{equation}
    \left\langle\text{tr}\left[\xi^{(\alpha)}_\tau\right]\text{tr}\left[\xi^{(\beta)}_\tau\right]\right\rangle = F_{\alpha\beta}(\vec{\phi}),
\end{equation}
where the expectation value is taken with respect to all the trajectories of time length $\tau$. 

We observe that the monitoring operators $\xi^{(\alpha)}_t$ can be evolved on quantum trajectories, in parallel with the state, with appropriate evolution equations~\cite{Albarelli2018}; in this way, the Fisher information can be computed on quantum trajectories. To make this computation even more efficient (specifically for small systems), we made use of the Fisher-Gillespie algorithm, introduced in~\cite{Radaelli2023}.

\bibliography{references.bib}

\end{document}